\documentclass[aps,prb,twocolumn,amsmath,amssymb,superscriptaddress,floatfix]{revtex4}
\usepackage{graphicx}
\usepackage{bm}
\bibstyle{apsrev.bib}

\newcommand{\vfi}{\varphi}
\newcommand{\vfr}{\varphi(\bm{r})}

\newcommand{\be}{\begin{equation}}
\newcommand{\ee}{\end{equation}}
\newcommand{\bea}{\begin{eqnarray}}
\newcommand{\eea}{\end{eqnarray}}
\newcommand{\pe}{{\scriptscriptstyle+}}
\newcommand{\me}{{\scriptscriptstyle-}}
\newcommand{\pme}{{\scriptscriptstyle\pm}}
\newcommand{\ksm}{K_{\scriptscriptstyle <}}
\newcommand{\kgr}{K_{\scriptscriptstyle >}}
\newcommand{\ssection}[2]{{$#1$}. {#2}.} 
\begin{document}

\title{Critical behavior at the interface between two systems belonging to different universality classes}

\author{Farkas \'A. Bagam\'ery}
 \affiliation{Institute of Theoretical Physics,
Szeged University, H-6720 Szeged, Hungary}
 \affiliation{Laboratoire de Physique des Mat\'eriaux, Universit\'e Henri 
Poincar\'e (Nancy 1), BP~239,\\
F-54506 Vand\oe uvre l\`es Nancy Cedex, France}
\author{Lo\"{\i}c Turban}
 \email{turban@lpm.uhp-nancy.fr}
 \affiliation{Laboratoire de Physique des Mat\'eriaux, Universit\'e Henri 
Poincar\'e (Nancy 1), BP~239,\\
F-54506 Vand\oe uvre l\`es Nancy Cedex, France}
\author{Ferenc Igl\'oi}%
 \email{igloi@szfki.hu}
 \affiliation{Research Institute for Solid State Physics and Optics,
H-1525 Budapest, P.O.Box 49, Hungary}
 \affiliation{Institute of Theoretical Physics,
Szeged University, H-6720 Szeged, Hungary}

\date{\today}

\begin{abstract}
We consider the critical behavior at an interface which separates two semi-infinite subsystems
belonging to different universality classes, thus having different set of critical exponents,
but having a common transition temperature. We solve this problem analytically in the frame of
$\phi^k$ mean-field theory, which is then generalized using phenomenological scaling considerations.
A large variety of interface critical behavior is obtained which is checked numerically on the
example of two-dimensional $q$-state Potts models with $2 \le q \le 4$. Weak interface couplings are generally  irrelevant, resulting in the same critical behavior at the interface as for a free surface. With strong interface couplings, the interface remains ordered at the bulk transition temperature. More interesting is the intermediate situation, the special interface transition, when the critical behavior at the interface involves new critical exponents, which however can be
expressed in terms of the bulk and surface exponents of the two subsystems. We discuss also the
smooth or discontinuous nature of the order parameter profile.

\end{abstract}


\maketitle

\section{Introduction}
\label{sec:intro}
Systems which undergo a second-order phase transition display singularities in different
physical observables which have been the subject of intensive research, both experimentally
and theoretically.\cite{dgl} At the critical temperature, $T_c$, due to the existence of a diverging correlation
length $\xi \sim |T-T_c|^{-\nu}$, 
microscopic inhomogeneities and single defects of finite size do not modify the critical
singularities which are observed in the perfect systems.\cite{burkhardt81_rev} However,
inhomogeneities of infinite extent,
such as the surface of the sample,\cite{binder83,diehl86,pleimling04} internal defect planes,\cite{burkhardt81}
etc., may modify the local critical
properties near the inhomogeneity, within a region with a characteristic size given by 
the correlation length. For example, the
magnetization, $m$, which vanishes in the bulk as $m \sim (T_c-T)^{\beta}$
behaves as $m_1 \sim (T_c-T)^{\beta_1}$ at a free surface\cite{binder83,diehl86,pleimling04} and the two critical exponents, $\beta$ and $\beta_1$,
are generally different.

Inhomogeneities having a more general form, such as localized\cite{bariev79} and extended
defects,\cite{hilhorst81} corners,\cite{corner}  wedges and edges, parabolic shapes,\cite{peschel91}  etc., often have exotic local critical behavior;
for a review, see Ref.~\onlinecite{ipt93}. The local critical behavior can be nonuniversal, so that the local
exponents vary  continuously with  some parameters, such as the opening angle of the corner,\cite{corner}
the amplitude of a localized,\cite{bariev79} or   extended defect.\cite{hilhorst81} The
inhomogeneity can also reduce the local order to such an extent that the local magnetization
vanishes with an essential singularity, as observed at the tip of a parabolic-shaped system.\cite{peschel91}  On the contrary, for enhanced local couplings, a surface or an interface may remain ordered at or above the bulk critical temperature,\cite{binder83,diehl86,pleimling04}  which in a two-dimensional (2D) system leads to a discontinuous local transition.\cite{it93}

In the problems we mentioned so far the inhomogeneities are embedded into a critical system the
bulk properties of which govern, among others, the divergence of the correlation length and the
behavior of the order-parameter profile. There is, however, another class of problems, in which
two (or more) systems meet at an interface,  each having different type of bulk (and
surface) critical properties. In this respect we can mention grain boundaries between
two different materials or the interface between two immiscible liquids, etc. 

If the critical temperatures of the two subsystems are largely different, the nature of the transitions at the interface is expected to be the same as for a surface.\cite{berche91} At the lower critical temperature, due to the presence of the nearby ordered  subsystem, the interface transition has the same properties as the extraordinary surface transition.\cite{binder83,diehl86,pleimling04}  At the upper critical temperature, the second subsystem being disordered, the interface transition is actually an ordinary surface transition.\cite{binder83,diehl86,pleimling04}  If the dimension of the system is larger than 2 and if the interface couplings are strong enough, one expects an interface transition in the presence of the two disordered subsystems whose properties should depend on the universality classes of these two subsystems.

Even in 2D, the local critical behavior at the interface can be more complex if the critical temperatures
of the subsystems are the same or if their difference is much smaller than the deviation from their mean value. In this case an interplay or competition between the two different bulk and surface critical behaviors can result in
a completely new type of interface critical phenomena. In this paper we study this
problem, assuming  that the critical temperatures of the two subsystems are identical. 

The structure of the paper is the following. The mean-field solution of the problem including $\varphi^3$,
$\varphi^4$ and $\varphi^6$  theories and the interface between them is presented in Sec.~\ref{sec:mf}. The mean-field results are generalized in Sec.~\ref{sec:scaling} using phenomenological scaling considerations. In Sec.~\ref{sec:num} these results are confronted with Monte Carlo simulations  in 2D for interfaces between subsystems belonging to the universality classes of the Ising model, the three-  and four-state Potts models, as well as the Baxter-Wu (BW) model.
Our results are discussed in Sec.~\ref{sec:disc} and some details about the analytical mean-field
calculations are given in the Appendixes A and B.

\section{Mean-field theory}
\label{sec:mf}
\subsection{Properties of the $\bm{\varphi^k}$ model}
\label{sec:mf-1}
\subsubsection{Free energy}
\label{sec:mf-1-1}
We consider a system with volume $V$ limited by a surface $S$ in the Landau mean-field approximation. The total free energy
is the sum of bulk and surface contributions which are functionals of the scalar order parameter $\vfr$ so that:\cite{binder83,diehl86,pleimling04}
\be
F[\vfi]=\int_{(V)}f_b\,[\vfi]\,dV + \int_{(S)} f_s\,[\vfi]\,dS \,.
\label{eq:F}
\ee
Near a second-order transition, the order parameter is small and the bulk free energy density $f_b[\vfi]$ is written as an expansion in the order parameter and its gradient, limited to the following terms:
\be
f_b\,[\vfi]=f_b\,[0]+\frac{C}{2}(\bm{\nabla}\vfi)^2+\frac{A}{2}\vfi^2+\frac{B}{k}\vfi^k-h\vfi\,.
\label{eq:fb}
\ee
The second term, with $C>0$, gives a positive contribution associated with the spatial variation of the order parameter. $A=-at$ ($a>0$, $t=T_c-T$) is negative when $T<T_c$ and measures the deviation from the critical point. The next term with $B>0$ ensures the stability of the system in the ordered phase. In the last term, $h$ is the bulk external field. When $k$ is odd, the order parameter is supposed to take only non-negative values; otherwise the system would be unstable. 

In the same way the surface free energy density is written as
\be
f_s\,[\vfi]=f_s\,[0]+\frac{C_s}{2}\frac{\vfi^2}{\Lambda}\,,
\label{eq:fs}
\ee
where $\vfi$ is the value of the order parameter on $(S)$. The constant $C_s$ is positive and $\Lambda$ is a characteristic length related to the surface and bulk couplings of the corresponding microscopic Hamiltonian of the system.\cite{binder83} 

\subsubsection{Ginzburg-Landau equation}
\label{sec:mf-1-2}
The mean-field equilibrium value of the order parameter, $\vfr$, minimizes the free energy in~(\ref{eq:F}). It is obtained through a variational method by calculating $\delta F[\vfi]$, the change of the free energy, which vanishes to first order in the deviation $\delta \vfr$ of the order parameter from its equilibrium value. Using Eqs.~(\ref{eq:F})--(\ref{eq:fs}), one obtains
\bea
\delta F[\vfi]&=&\int_{(V)}[C\bm{\nabla}\vfi\cdot\bm{\nabla}\delta\vfi+(A\vfi+B\vfi^{k-1}-h)\delta\vfi]\,dV\nonumber\\
&&\ \ \ \ \ + \int_{(S)}\left(C_s\frac{\vfi}{\Lambda}\right)\delta\vfi\, dS\,.
\label{eq:deltaF1}
\eea
The first term in the volume integral may be rewritten as
\be
C\bm{\nabla}\vfi\cdot\bm{\nabla}\delta\vfi=\bm{\nabla}\cdot(C\delta\vfi\bm{\nabla}\vfi)-C\delta\vfi\nabla^2\vfi\,,
\label{eq:ident}
\ee
and the contribution to~(\ref{eq:deltaF1}) of the first term on the right can be transformed into a surface integral through Gauss' theorem. Then
\bea
\delta F[\vfi]&=&\int_{(V)}(-C{\nabla}^2\vfi+A\vfi+B\vfi^{k-1}-h)\delta\vfi\,dV\nonumber\\
&&\ \ \ \ \ + \int_{(S)}\left(-C\bm{n}\cdot\bm{\nabla}\vfi+C_s\frac{\vfi}{\Lambda}\right)\delta\vfi\, dS\,,
\label{eq:deltaF2}
\eea
where $\bm{n}$ is  a unit vector normal to the surface and pointing inside the system.

At equilibrium, the first-order variation of the free energy vanishes. The volume integral leads to the Ginzburg-Landau equation
\be
-C{\nabla}^2\vfr+A\vfr+B\vfi^{k-1}(\bm{r})=h(\bm{r})
\label{eq:GL}
\ee
governing the equilibrium behavior of the order parameter in the volume of the system and the surface integral provides the boundary condition: 
\be
C\,\bm{n}\cdot\bm{\nabla}\vfr\vert_{(S)}=C_s\left.\frac{\vfr}{\Lambda}\right\vert_{(S)}\,.
\label{eq:bc}
\ee

\subsubsection{Bulk critical behavior}
\label{sec:mf-1-3}
In the bulk, the first term in~(\ref{eq:GL}) vanishes. The zero-field magnetization vanishes when $T>T_c$ and is given by
\be
\vfi_b=\left(\frac{at}{B}\right)^\beta=\vfi_0\,t^\beta\,,\qquad \beta=\frac{1}{k-2}\,,
\label{eq:mb}
\ee
in the ordered phase ($T\leq T_c$, $h=0$).

The connected part of order-parameter two-point correlation function is given by
\be
G(\bm{r},\bm{r}')=k_{\rm B}T\frac{\delta\vfr}{\delta h(\bm{r}')}
\label{eq:cor}
\ee
where $\vfr$ is the equilibrium order parameter, solution of the Ginzburg-Landau equation. Taking the functional derivative of Eq.~(\ref{eq:GL}), one obtains:
\be
-C\nabla^2_{\bm{r}}G(\bm{r},\bm{r}')+[A+(k-1)B\vfi_b^{k-2}]G(\bm{r},\bm{r}')=k_{\rm B}T\delta(\bm{r}-\bm{r}')\,.
\label{eq:fd1}
\ee
This may be rewritten as 
\be
(-\nabla^2_{\bm{r}}+\xi^{-2})G(\bm{r},\bm{r}')=\frac{k_{\rm B}T}{C}\delta(\bm{r}-\bm{r}')\,,
\label{eq:fd2}
\ee
where the expression of the bulk correlation length $\xi$ follows from Eqs.~(\ref{eq:mb}) and~(\ref{eq:fd1}) and reads
\be
\xi=\left[(k-2)\frac{at}{C}\right]^{-\nu}=\xi_0\,t^{-\nu}\,,\qquad \nu=\frac{1}{2}\,,
\label{eq:xi}
\ee
in the ordered phase.

\subsubsection{Order parameter profiles}
\label{sec:mf-1-4}
We now assume that the surface of the system is located at $z=0$ so that $\vfi=\vfi(z)$. 
Then, according to Eqs.~(\ref{eq:GL}), (\ref{eq:mb}) and (\ref{eq:xi}), the zero-field normalized order parameter profile $\hat{\vfi}=\vfi/\vfi_b$ is the solution of the following differential equation:
\be
\frac{d^2\hat{\vfi}}{dz^2}=\frac{A}{C}\,\hat{\vfi}+\frac{B}{C}\,\vfi_b^{k-2}\hat{\vfi}^{k-1}
=\frac{\hat{\vfi}^{k-1}-\hat{\vfi}}{(k-2)\,\xi^2}\,.
\label{eq:dif1}
\ee
Multiplying by $2D\hat{\vfi}/dz$ and taking into account the bulk boundary condition, $d\hat{\vfi}/dz\to0$ when $\hat{\vfi}\to1$, a first integration leads to:
\be
\left(\frac{d\hat{\vfi}}{dz}\right)^2=\frac{2\hat{\vfi}^k-k\hat{\vfi}^2+k-2}{k(k-2)\,\xi^2}\,.
\label{eq:dif2}
\ee
To go further we have to specify the value of $k$ and to distinguish between surfaces (or interfaces) which are more ordered ($\hat{\vfi}(0)>1$) or less ordered ($\hat{\vfi}(0)<1$) than the bulk. Below  we list the solutions of Eq.~(\ref{eq:dif2}) which will be needed in the sequel. We use the notation $\hat{\vfi}_\pe(z)$ [$\hat{\vfi}_\me(z)$] for a system located in the $z>0$ [$z<0$]  half-space. The values of the integration constants $l_\pe$ and $l_\me$ are determined by the boundary conditions at $z=0$.

$\vfi^3$ model, $\hat{\vfi}(0)>1$:
\bea
\hat{\vfi}_\pme (z)&=&\frac{1}{2}\left[3\coth^2\left(\frac{z\pm l_\pme}{2\xi_\pme}\right)-1\right]\,,\nonumber\\
\left.\frac{d\hat{\vfi}_\pme}{dz}\right\vert_0&=&\mp\frac{3}{2\xi_\pme}\cosh\left(\frac{l_\pme}{2\xi_\pme}\right)
\sinh^{-3}\left(\frac{l_\pme}{2\xi_\pme}\right)\,.
\label{eq:fi3-a}
\eea

$\vfi^3$ model, $\hat{\vfi}(0)<1$:
\bea
\hat{\vfi}_\pme (z)&=&\frac{1}{2}\left[3\tanh^2\left(\frac{z\pm l_\pme}{2\xi_\pme}\right)-1\right]\,,\nonumber\\
\left.\frac{d\hat{\vfi}_\pme}{dz}\right\vert_0&=&\pm\frac{3}{2\xi_\pme}\sinh\left(\frac{l_\pme}{2\xi_\pme}\right)
\cosh^{-3}\left(\frac{l_\pme}{2\xi_\pme}\right)\,.
\label{eq:fi3-b}
\eea

$\vfi^4$ model, $\hat{\vfi}(0)>1$:
\bea
\hat{\vfi}_\pme (z)&=&\pm\coth\left(\frac{z\pm l_\pme}{2\xi_\pme}\right)\,,\nonumber\\
\left.\frac{d\hat{\vfi}_\pme}{dz}\right\vert_0&=&\mp\frac{1}{2\xi_\pme}
\sinh^{-2}\left(\frac{l_\pme}{2\xi_\pme}\right)\,.
\label{eq:fi4-a}
\eea

$\vfi^4$ model, $\hat{\vfi}(0)<1$:
\bea
\hat{\vfi}_\pme(z)&=&\pm\tanh\left(\frac{z\pm l_\pme}{2\xi_\pme}\right)\,,\nonumber\\
\left.\frac{d\hat{\vfi}_\pme}{dz}\right\vert_0&=&\pm\frac{1}{2\xi_\pme}
\cosh^{-2}\left(\frac{l_\pme}{2\xi_\pme}\right)\,.
\label{eq:fi4-b}
\eea

$\vfi^6$ model, $\hat{\vfi}(0)>1$:
\bea
\hat{\vfi}_\pme(z)&=&\sqrt{2}\left[3\tanh^2\left(\frac{z\pm l_\pme}{2\xi_\pme}\right)-1\right]^{-1/2}\!\!\!\!\!\!\!\!\!\,,\nonumber\\
\left.\frac{d\hat{\vfi}_\pme}{dz}\right\vert_0&=&\mp\frac{3\sqrt{2}}{2\xi_\pme}\sinh\left(\frac{l_\pme}{2\xi_\pme}\right)
\cosh^{-3}\left(\frac{l_\pme}{2\xi_\pme}\right)\nonumber\\
&&\ \ \times\left[3\tanh^2\left(\frac{l_\pme}{2\xi_\pme}\right)-1\right]^{-3/2}\!\!\!\!\!\!\!\!\!\,.
\label{eq:fi6-a}
\eea

$\vfi^6$ model, $\hat{\vfi}(0)<1$:
\bea
\hat{\vfi}_\pme(z)&=&\pm\sinh\left(\frac{z\pm l_\pme}{2\xi_\pme}\right)
\left[\sinh^2\left(\frac{z\pm l_\pme}{2\xi_\pme}\right)+\frac{3}{2}\right]^{-1/2}\!\!\!\!\!\!\!\!\!\,,\nonumber\\
\left.\frac{d\hat{\vfi}_\pme}{dz}\right\vert_0\!\!&=&\!\pm\frac{3}{4\xi_\pme}\cosh\left(\frac{l_\pme}{2\xi_\pme}\right)
\left[\sinh^2\left(\frac{l_\pme}{2\xi_\pme}\right)+\frac{3}{2}\right]^{-3/2}\!\!\!\!\!\!\!\!\!\!\!\!\,.
\label{eq:fi6-b}
\eea

\subsection{Surface critical behavior}
\label{sec:mf-2}

In this section we briefly consider, for later use, the critical behavior at the surface when the bulk is in its ordered phase ($T\leq T_c$). 
We suppose that the system is located in the $z>0$ half-space. Since there is no ambiguity here we drop 
the index $+$ so that, for example, $\vfi(z)$ stands for $\vfi_\pe(z)$. 
In this geometry, according to Eq.~(\ref{eq:bc}), the boundary condition reads
\be
C_s\frac{\hat{\vfi}(0)}{\Lambda}=C\left.\frac{d\hat{\vfi}}{dz}\right\vert_0\,.
\label{eq:bcz}
\ee
Below we list the values obtained for the integration constant $l$ and the surface order parameter $\vfi(0)$. 

\subsubsection{$\vfi^3$ surface}
\label{sec:mf-2-1}
\ssection{a}{$\Lambda<0$}
When $\Lambda<0$, the surface remains ordered at the bulk critical point, which corresponds to the {\it extraordinary surface transition}. Thus, the order parameter profile is given by Eq.~(\ref{eq:fi3-a}) and $l$ is obtained by expanding both sides of Eq.~(\ref{eq:bcz}) in powers of $l/\xi\ll1$. To leading order, one obtains
\be
l=2\frac{C}{C_s}\vert\Lambda\vert
\label{eq:l3-a}
\ee
and
\be
\vfi(0)=\frac{3}{2}\left(\frac{C_s\xi_0}{C\vert\Lambda\vert}\right)^2\vfi_0\,.
\label{eq:f3-a}
\ee

\ssection{b}{$\Lambda>0$}
In this case the surface is less ordered than the bulk and we have an {\it ordinary surface transition}. The profile is given by Eq.~(\ref{eq:fi3-b}) and the solution is obtained by assuming that the ratio $l/\xi$ is a constant. Details of the calculation are given in Appendix~\ref{app:surf}. The boundary condition in~(\ref{eq:bcz}) is satisfied with
\be
l=2\tanh^{-1}\left(\frac{1}{\sqrt{3}}\right)\xi_0 t^{-1/2}
\label{eq:l3-b}
\ee
so that:
\be
\vfi(0)=\frac{1}{\sqrt{3}}\frac{C\Lambda}{C_s\xi_0}\vfi_0 t^{3/2}\,.
\label{eq:f3-b}
\ee
Thus the surface exponent, $\beta_1=3/2$, is larger than the bulk exponent, $\beta=1$.

\ssection{c}{$\Lambda\to\infty$}
The profile is given by either~(\ref{eq:fi3-a}) or~(\ref{eq:fi3-b}) with $l\to\infty$. Then
the order parameter is constant keeps its bulk value until the surface and $\vfi(0)=\vfi_b=\vfi_0t$. We have a
{\it special surface transition}, which corresponds to a multicritical point where the lines of ordinary and extraordinary transitions  meet with the line of surface transition\cite{binder83} in a ($T,1/\Lambda$) diagram.

\subsubsection{$\vfi^4$ surface}
\label{sec:mf-2-2}

\ssection{a}{$\Lambda<0$}
This corresponds as above to the extraordinary transition where the surface remains ordered at the bulk critical point. The profile is given by Eq.~(\ref{eq:fi4-a}) and the boundary condition in~(\ref{eq:bcz}) requires $l/\xi\ll1$ so that one obtains
\be
l=\frac{C}{C_s}\vert\Lambda\vert\,.
\label{eq:l4-a}
\ee
The leading contribution to the surface order parameter is given by
\be
\vfi(0)=2\frac{C_s\xi_0}{C\vert\Lambda\vert}\vfi_0\,.
\label{eq:f4-a}
\ee

\ssection{b}{$\Lambda>0$}
At the ordinary surface transition, the profile is given by Eq.~(\ref{eq:fi4-b}). Here the boundary condition is satisfied with $l/\xi\ll1$,
which gives
\be
l=\frac{C}{C_s}\Lambda\,.
\label{eq:l4-b}
\ee
The surface order parameter vanishes as
\be
\vfi(0)=\frac{1}{2}\frac{C\Lambda}{C_s\xi_0}\vfi_0 t\,.
\label{eq:f4-b}
\ee
Thus, the surface exponent is $\beta_1=1$ to be compared to the bulk exponent, $\beta=1/2$.

\ssection{c}{$\Lambda\to\infty$}
Here too, the boundary condition  in~(\ref{eq:bcz}) leads to $l\to\infty$ and the surface order parameter keeps the bulk value, $\vfi(0)=\vfi_0t^{1/2}$, at the special surface transition.

\subsubsection{$\vfi^6$ surface}
\label{sec:mf-2-3}

\ssection{a}{$\Lambda<0$}
Once more we have an extraordinary surface transition with a profile given by Eq.~(\ref{eq:fi6-a}). As shown in Appendix~\ref{app:surf} this is another instance where the boundary condition in~(\ref{eq:bcz}) is satisfied 
with a constant value of the ratio $l/\xi$. Thus, as in (\ref{eq:l3-b}), we have
\be
l=2\tanh^{-1}\left(\frac{1}{\sqrt{3}}\right)\xi_0 t^{-1/2}
\label{eq:l6-a}
\ee
The leading contribution to the surface order parameter is then
\be
\vfi(0)=\left(2\sqrt{3}\frac{C_s\xi_0}{C\vert\Lambda\vert}\right)^{1/2}\vfi_0\,.
\label{eq:f6-a}
\ee

\ssection{b}{$\Lambda>0$}
The profile at the ordinary surface transition is given by Eq.~(\ref{eq:fi6-b}). Here we have the standard behavior, $l/\xi\ll1$, with
\be
l=\frac{C}{C_s}\Lambda\,.
\label{eq:l6-b}
\ee
The surface order parameter displays the following behavior:
\be
\vfi(0)=\frac{1}{\sqrt{6}}\frac{C\Lambda}{C_s\xi_0}\vfi_0 t^{3/4}\,.
\label{eq:f6-b}
\ee
Thus the surface exponent is $\beta_1=3/4$ whereas $\beta=1/4$ in the bulk.

\ssection{c}{$\Lambda\to\infty$}

As for the other models, the length $l$ is infinite at the special transition and the surface order parameter has the bulk value, $\vfi(0)=\vfi_0t^{1/4}$.

Although the characteristic length $l$ sometimes remains finite and sometimes diverges at the critical point, the exponent $\beta_1$ at the ordinary surface transition always satisfies the scaling relation:
\be
\beta_1=\beta+\nu\qquad(\mathrm{ordinary\  transition})\,.
\label{eq:scal1}
\ee
In the same way, at the special transition, we have:
\be
\beta_1=\beta\qquad(\mathrm{special\  transition})\,.
\label{eq:scal2}
\ee
One should notice that these scaling relations are only valid in mean-field theory.\cite{binder83}

At the extraordinary transition the singular term, governing the approach to the constant value at $T_c$ of the surface order parameter, appears at the next order in the expansion. It vanishes linearly in $t$ for the $\vfi^3$ and $\vfi^4$ models and as $t^{1/2}$ for the $\vfi^6$ model. We do not give further details since we shall not need it in the following. For the same reason, we did not examine the properties of the surface transition which occurs in the surface region, above the bulk critical temperature, when $\Lambda<0$.

\subsection{Interface critical behavior}
\label{sec:mf-3}
In this section, we consider the critical behavior at the interface  between two systems, belonging to different universality classes, in their ordered phase ($T\leq T_c$). Thus the free energy densities of the two subsystems are given by~(\ref{eq:fb}) with different values of $k$. They are coupled through an interface at $z=0$ with free energy density
\be
f_i\,[\vfi]=f_i\,[0]+\frac{C_i}{2}\frac{\vfi^2}{\Lambda}\,.
\label{eq:fi}
\ee
We assume that the positive half-space corresponds to the system which is the more ordered in the bulk when $T_c$ is approached from below so that $\beta^\pe<\beta^\me$. The order parameter profiles, $\vfi_\pe(z)$ for $z>0$ and $\vfi_\me(z)$ for $z<0$, have now to satisfy
\bea
\vfi(0)&=&\vfi_\pe(0)=\vfi_\me(0)\,,\nonumber\\
C_i\frac{\vfi(0)}{\Lambda}&=&C\!_\pe\left.\frac{d\vfi_\pe}{dz}\right\vert_0-C\!_\me\left.\frac{d\vfi_\me}{dz}\right\vert_0\,.
\label{eq:bci}
\eea
These  boundary conditions generalize Eq.~(\ref{eq:bcz}) for the interface geometry where each subsystem contributes a normal derivative to the surface integral in Eq.~(\ref{eq:deltaF2}). 

When two subsystems are coupled, the boundary conditions in Eq.~(\ref{eq:bci})  determine the integration constants $l_\pme$---i.e., the complete order parameter profile.
In the following, we give these integration constants as well as $\vfi(0)$, the value of order parameter at the interface, for the different types of interface considered. Technical details about the calculations can be found in Appendix~\ref{app:inter}.

As in the surface case, depending on the value of $\Lambda$, different types of interface critical behaviors are obtained (see Figs.~1--3). When $\Lambda<0$, the interface remains ordered at the bulk critical point and we have an {\it extraordinary interface transition}. When $d>2$,  the local order persists above the bulk critical temperature until a $\Lambda$-dependant {\it interface transition} temperature is reached. This transition, which always occurs in mean-field theory, will not be discussed further here. When $\Lambda>0$ the interface order parameter vanishes at the bulk $T_c$ as a power of $t$. This corresponds to the {\it interface ordinary transition}.  When parameterized by $1/\Lambda$, these two transition lines meet, together with the interface transition line when it exists, at a multicritical point corresponding to the {\it special interface transition} located at $1/\Lambda=0,T=T_c$.

\subsubsection{$\vfi^3-\vfi^4$ interface}
\label{sec:mf-3-1}

\ssection{a}{$\Lambda<0$}
This corresponds to strong couplings at the interface. The order parameter increases when the interface is approached so that $\vfi_\me(z)$ and $\vfi_\pe(z)$ are given by~(\ref{eq:fi3-a}) and~(\ref{eq:fi4-a}), respectively. To leading order in $t$, we have 
\bea
l_\me&=&f\frac{C\!_\me}{C_i} \vert\Lambda\vert\,,\nonumber\\
l_\pe&=&\frac{1}{3}\frac{\vfi_0^\pe}{\vfi_0^\me}\left(f\frac{C\!_\me \Lambda}{C_i\xi_0^\me}\right)^2\xi_0^\pe\,,
\label{eq:z34-a}
\eea
where
\be
f=1+\sqrt{1+3\frac{\Lambda^*}{\vert\Lambda\vert}}\,,\qquad  
 \Lambda^*=\frac{C\!_\pe C_i\vfi_0^\me (\xi_0^\me)^2}{C_\me^2\vfi_0^\pe\xi_0^\pe}\,.
\label{eq:f34-a}
\ee
The leading contribution to the order parameter at the interface,
\be
\vfi(0)=6\left(\frac{1}{f}\frac{C_i\xi_0^\me}{C\!_\me \Lambda} \right)^2\vfi_0^\me\,,
\label{eq:fi34-a}
\ee
is also independent of $t$; i.e.,  the interface remains ordered at the bulk critical point. 

According to~(\ref{eq:f34-a}) and~(\ref{eq:fi34-a}), the asymptotic dependence on $\vert\Lambda\vert$ is the following:
\be
\vfi(0)\propto\left\{
\begin{array}{ll}
\vert\Lambda\vert^{-2}&\vert\Lambda\vert\gg\Lambda^*\\
\vert\Lambda\vert^{-1}&\vert\Lambda\vert\ll\Lambda^*\\
\end{array}
\right.
\label{eq:la34}
\ee

\ssection{b}{$\Lambda>0$}
This corresponds to weak couplings between the two subsystems. 
When 
\be
0<\Lambda<\Lambda_c=2\frac{C_i \vfi_0^\me}{C\!_\pe\vfi_0^\pe}\xi_0^\pe\,,
\label{eq:lc}
\ee
the order parameter decreases from both sides towards the interface. Then $\vfi_\me(z)$ is given by~(\ref{eq:fi3-b}) and $\vfi_\pe(z)$ by~(\ref{eq:fi4-b}) with
\bea
l_\me&=&2 \tanh^{-1}\left(\sqrt{\frac{1+2\Lambda/\Lambda_c}{3}}\right)\xi_0^\me  t^{-1/2}\,,\nonumber\\
l_\pe&=&\frac{C\!_\pe}{C_i}\Lambda\,.
\label{eq:z34-b1}
\eea

$l_\me$ diverges when $\Lambda=\Lambda_c$. Then $\vfi_\me(z)$ is a constant and keeps its bulk value for any $z\leq0$. 

When $\Lambda_c<\Lambda<\infty$, $\vfi_\pe(z)$ is still given by~(\ref{eq:fi4-b}) and $l_\pe$ keeps the value given in Eq.~(\ref{eq:z34-b1}) but now $\vfi_\me(z)$, which increases, is given by Eq.~(\ref{eq:fi3-a}) with
\be
l_\me=2\coth^{-1}\left(\sqrt{\frac{1+2\Lambda/\Lambda_c}{3}}\right)\xi_0^\me t^{-1/2}\,.
\label{eq:z34-b2}
\ee
For $0<\Lambda<\infty$, the order parameter at the interface is always given by
\be
\vfi(0)=\frac{1}{2}\frac{C\!_\pe\Lambda}{C_i\xi_0^\pe}\vfi_0^\pe t\,.
\label{eq:fi34-b}
\ee


\begin{figure}[tbh]
\includegraphics[width=\columnwidth]{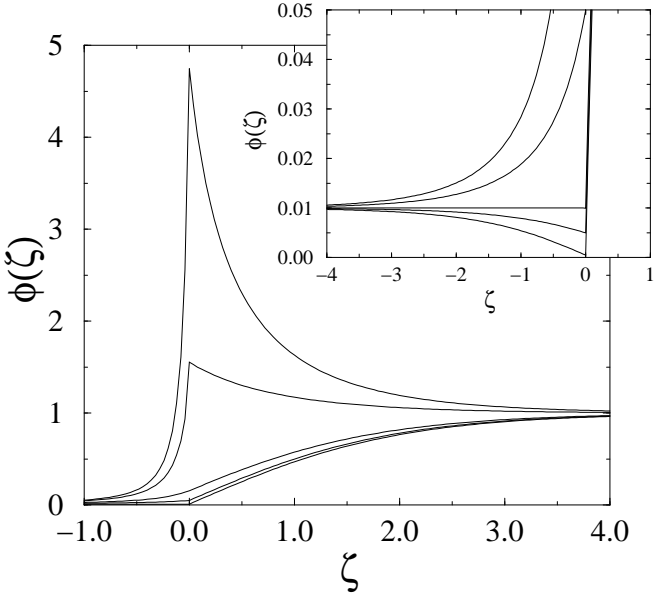}
\caption{Reduced order parameter profile $\phi=\vfi(z)/\vfi_b^\pe$ at the $\vfi^3-\vfi^4$ interface as a function of $\zeta=z/\xi_\pe$ for different values of $\lambda=\Lambda/\xi_0^\pe$ with $1/\lambda=-0.2$, $-0.1$, 0, $0.1$ and $0.5$, from top to bottom.  The behavior for $\lambda\geq0$ and $\zeta<0$ is enlarged in the inset where $1/\lambda=0$, $0.1$, $0.5$, $1$ and $10$, from top to bottom. One may notice the change of behavior at $1/\lambda_c=\xi_0^\pe/\Lambda_c=1/2$. All other parameters have the same values in the two subsystems: $C\!_\pe=C\!_\me=C_i$, $\vfi_0^\pe=\vfi_0^\me$, $\xi_0^\pe=\xi_0^\me$ and $t=10^{-4}$.}
\label{fig1}
\end{figure}



\begin{figure}[tbh]
\includegraphics[width=\columnwidth]{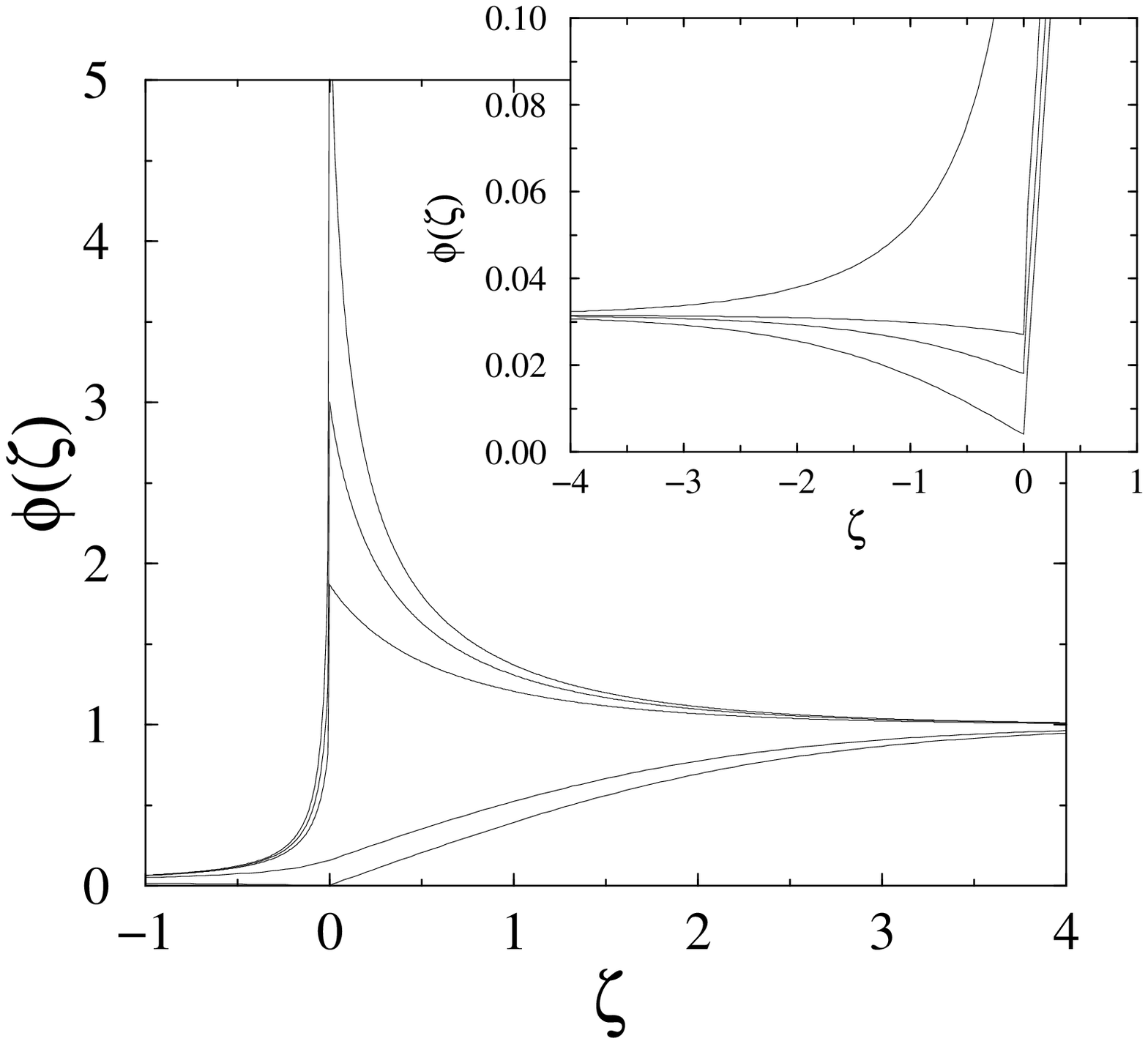}
\caption{Reduced order parameter profile $\phi=\vfi(z)/\vfi_b^\pe$ at the $\vfi^4-\vfi^6$ interface as a function of $\zeta=z/\xi_\pe$ for different values of $\lambda=\Lambda/\xi_0^\pe$ with $1/\lambda=-0.1$, $-0.05$, $-0.03$, $0$ and $0.1$, from top to bottom.  The behavior for $\lambda\geq0$ and $\zeta<0$ is enlarged in the inset where $1/\lambda=0$, $0.01$, $0.02$ and $0.1$, from top to bottom. All other parameters have the same values in the two subsystems and $t=10^{-6}$.}
\label{fig2}
\end{figure}


\ssection{c}{$\Lambda\to\infty$}
The profile remains monotonously increasing and keeps the same functional form as for $\Lambda>\Lambda_c$ although $l_\me$ and $l_\pe$ are now given by
\bea
l_\me&=&2\left(3\frac{C\!_\me\vfi_0^\me}{C\!_\pe\vfi_0^\pe}\xi_0^\pe\right)^{1/3}(\xi_0^\me)^{2/3}t^{-1/3}\,,\nonumber\\
l_\pe&=&\left(3\frac{\vfi_0^\me}{\vfi_0^\pe}\xi_0^\pe\right)^{1/3}
\left(\frac{C\!_\pe}{C\!_\me}\xi_0^\me\right)^{2/3}t^{-1/3}\,,
\label{eq:z34-c}
\eea
so that
\be
\vfi(0)=\frac{1}{2}\left(3\vfi_0^\me\right)^{1/3}
\left(\frac{C\!_\pe\xi_0^\me}{C\!_\me\xi_0^\pe}\vfi_0^\pe\right)^{2/3}t^{2/3}\,.
\label{eq:fi34-c}
\ee

\subsubsection{$\vfi^4-\vfi^6$ interface}
\label{sec:mf-3-2}

\ssection{a}{$\Lambda<0$}
The interface is more ordered than the bulk. Thus $\vfi_\me(z)$ is given by~(\ref{eq:fi4-a}) and $\vfi_\pe(z)$ by~(\ref{eq:fi6-a}) with
\bea
l_\me&=&\frac{f}{2}\frac{C\!_\me}{C_i}\vert\Lambda\vert\,,\nonumber\\
l_\pe&=&2\left[\tanh^{-1}\left(\frac{1}{\sqrt{3}}\right)+h(t)\right]\xi_0^\pe t^{-1/2}\,,\nonumber\\
h(t)&=&\frac{\sqrt{3}}{32}\left(\frac{C\!_\me\vfi_0^\pe\vert\Lambda\vert}{C_i\vfi_0^\me\xi_0^\me}f\right)^2t^{1/2}\,,
\label{eq:z46-a}
\eea
where
\be
f=1+\sqrt{1+\frac{8}{\sqrt{3}}\frac{\Lambda^*}{\vert\Lambda\vert}}\,,\quad 
\Lambda^*=\frac{C_iC\!_\pe(\vfi_0^\me\xi_0^\me)^2}{(C\!_\me\vfi_0^\pe)^2\xi_0^\pe}
\label{eq:f46-a}
\ee
Here and below in Eq.~(\ref{eq:z36-a}) we keep the next-to-leading term $h(t)$ in $l_\pe$.
This correction is actually needed to obtain the correct form of the profile in the vicinity of $z=0$.
The leading contribution to the order parameter at the interface is constant:
\be
\vfi(0)=\frac{4}{f}\frac{C_i\xi_0^\me}{C\!_\me\vert\Lambda\vert}\vfi_0^\me\,.
\label{eq:fi46-a}
\ee
Its asymptotic behavior
\be
\vfi(0)\propto\left\{
\begin{array}{ll}
\vert\Lambda\vert^{-1}&\vert\Lambda\vert\gg\Lambda^*\\
\vert\Lambda\vert^{-1/2}&\vert\Lambda\vert\ll\Lambda^*\\
\end{array}
\right.\,,
\label{eq:la46}
\ee
follows from Eqs.~(\ref{eq:f46-a}) and~(\ref{eq:fi46-a})

\ssection{b}{$\Lambda>0$}
The profile is always decreasing when the interface is approached. Thus $\vfi_\me(z)$ is given by~(\ref{eq:fi4-b}) and $\vfi_\pe(z)$ by~(\ref{eq:fi6-b}) with the following expressions for the integration constants:
\bea
l_\me&=&\sqrt{\frac{2}{3}}\frac{C\!_\pe\vfi_0^\pe\xi_0^\me}{C_i\vfi_0^\me\xi_0^\pe}\Lambda t^{-1/4}\,,\nonumber\\
l_\pe&=&\frac{C\!_\pe}{C_i}\Lambda\,.
\label{eq:z46-b}
\eea
The interface order parameter behaves as
\be
\vfi(0)=\frac{1}{\sqrt{6}}\frac{C\!_\pe\Lambda}{C_i\xi_0^\pe}\vfi_0^\pe t^{3/4}\,.
\label{eq:fi46-b}
\ee

\ssection{c}{$\Lambda\to\infty$}

Then the profile increases monotonously with $z$. $\vfi_\me(z)$ is given by~(\ref{eq:fi4-a}) and $\vfi_\pe(z)$ by~(\ref{eq:fi6-b}) with
\bea
l_\me&=&\left(2\sqrt{6}\frac{C\!_\me\vfi_0^\me}{C\!_\pe\vfi_0^\pe}\xi_0^\me\xi_0^\pe\right)^{1/2}t^{-3/8}\,,\nonumber\\
l_\pe&=&\left(2\sqrt{6}\frac{C\!_\pe\vfi_0^\me}{C\!_\me\vfi_0^\pe}\xi_0^\me\xi_0^\pe\right)^{1/2}t^{-3/8}\,.
\label{eq:z46-c}
\eea
The interface order parameter vanishes as
\be
\vfi(0)=\left(\sqrt{\frac{2}{3}}\frac{C\!_\pe\xi_0^\me}{C\!_\me\xi_0^\pe}\vfi_0^\me\vfi_0^\pe\right)^{1/2}t^{3/8}\,.
\label{eq:fi46-c}
\ee

\subsubsection{$\vfi^3-\vfi^6$ interface}
\label{sec:mf-3-3}

\ssection{a}{$\Lambda<0$}
As usual in this case, the interface is more ordered than the bulk. The profiles, $\vfi_\me(z)$ and $\vfi_\pe(z)$ are given 
by Eqs.~(\ref{eq:fi3-a}) and~(\ref{eq:fi6-a}). The calculation of $l_\me$ involves the solution of an equation of the fourth degree (see Appendix~\ref{app:inter}). Here we only report the limiting behavior for large and small values 
of~$\vert\Lambda\vert$:
\bea
l_\me&\sim&2\frac{C\!_\me}{C_i}\vert\Lambda\vert\,,\qquad \vert\Lambda\vert\gg\Lambda^*\,,\nonumber\\
l_\me&\sim&\!\!\left(2\sqrt{27}\frac{C\!_\pe\vert\Lambda\vert}{C_i\xi_0^\pe}\right)^{1/4}
\!\!\!\left(\frac{\vfi_0^\me}{\vfi_0^\pe}\right)^{1/2}\!\!\!\!\xi_0^\me\,,\quad\vert\Lambda\vert\ll\Lambda^*\,,\nonumber\\
l_\pe&=&2\left[\tanh^{-1}\left(\frac{1}{\sqrt{3}}\right)+h(t)\right]\xi_0^\pe t^{-1/2}\,,\nonumber\\
h(t)&=&\frac{2\sqrt{3}}{9}\left(\frac{C\!_\me\vert\Lambda\vert}{C_i\xi_0^\me}\right)^4
\left(\frac{\vfi_0^\pe}{\vfi_0^\me}\right)^2t^{1/2}\,,\quad \vert\Lambda\vert\gg\Lambda^*\,,\nonumber\\
h(t)&=&\frac{1}{4}\frac{C\!_\pe\vert\Lambda\vert}{C_i\xi_0^\pe}\,t^{1/2}\,,\qquad \vert\Lambda\vert\ll\Lambda^*\,.
\label{eq:z36-a}
\eea
The crossover is taking place around
\be
\Lambda^*=C_i
\left(\frac{\vfi_0^\me}{\vfi_0^\pe}\right)^{2/3}
\left(\frac{C\!_\pe}{\xi_0^\pe}\right)^{1/3}
\left(\frac{\xi_0^\me}{C\!_\me}\right)^{4/3}\,.
\label{eq:cross}
\ee
The interface order parameter reads
\bea
\vfi(0)&\sim&\frac{3}{2}\left(\frac{C_i\xi_0^\me}{C\!_\me\Lambda}\right)^2\vfi_0^\me   \,,\qquad \vert\Lambda\vert\gg\Lambda^*\,,\nonumber\\
\vfi(0)&\sim&\left(2\sqrt{3}\frac{C_i\xi_0^\pe}{C\!_\pe\vert\Lambda\vert}\right)^{1/2}\vfi_0^\pe   \,,\qquad \vert\Lambda\vert\ll\Lambda^*\,,
\label{eq:fi36-a}
\eea


\begin{figure}[tbh]
\includegraphics[width=\columnwidth]{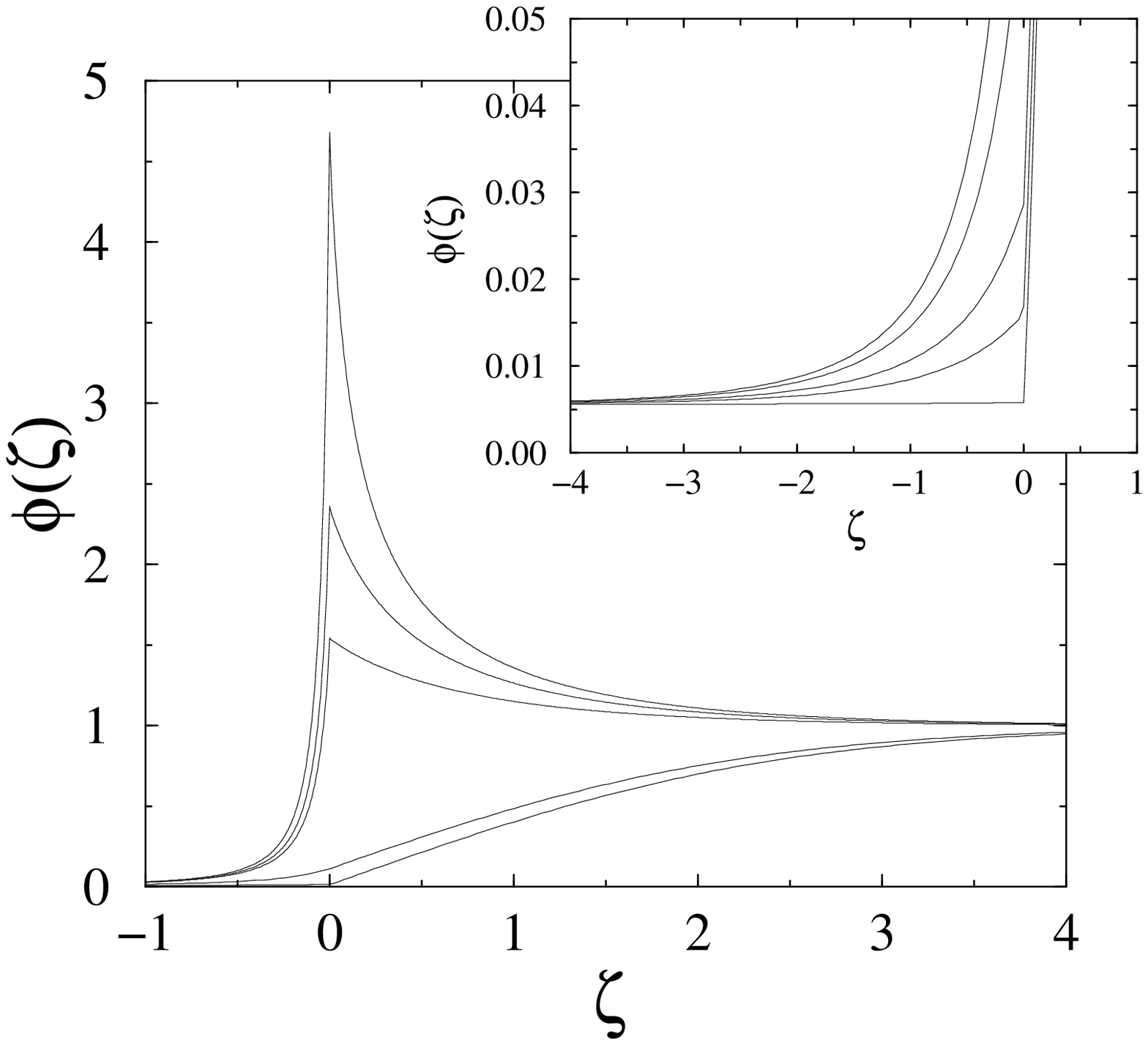}
\caption{Reduced order parameter profile $\phi=\vfi(z)/\vfi_b^\pe$ at the $\vfi^3-\vfi^6$ interface as a function of $\zeta=z/\xi_\pe$ for different values of $\lambda=\Lambda/\xi_0^\pe$ with $1/\lambda=-0.2$, $-0.05$, $-0.02$, $0$ and $1$, from top to bottom.  The behavior for $\lambda\geq0$ and $\zeta<0$ is enlarged in the inset where $1/\lambda=0$, $0.2$, $0.5$, $1$ and $10$, from top to bottom. All other parameters have the same values in the two subsystems and $t=10^{-3}$.}
\label{fig3}
\end{figure}


\ssection{b}{$\Lambda>0$}
The profile is monotonously increasing. $ \vfi_\me(z)$ and $ \vfi_\pe(z)$ have the form given in Eqs.~(\ref{eq:fi3-a}) 
and~(\ref{eq:fi6-b}) with the following values of the constants:
\bea
l_\me&=&\left(6\sqrt{6}\frac{C_i\vfi_0^\me\xi_0^\pe}{C\!_\pe\vfi_0^\pe\Lambda}\right)^{1/2}\xi_0^\me t^{-3/8}\,,\nonumber\\
l_\pe&=&\frac{C\!_\pe}{C_i}\Lambda\,.
\label{eq:z36-b}
\eea
The interface order parameter vanishes as
\be
\vfi(0)=\frac{1}{\sqrt{6}}\frac{C\!_\pe\Lambda}{C_i\xi_0^\pe}\vfi_0^\pe t^{3/4}\,.
\label{eq:fi36-b}
\ee

\ssection{c}{$\Lambda\to\infty$}
The profile is still given by Eqs.~(\ref{eq:fi3-a}) and~(\ref{eq:fi6-b}) with the following values of the constants:
\bea
l_\me&=&\left[12\sqrt{6}\frac{C\!_\me\vfi_0^\me}{C\!_\pe\vfi_0^\pe}(\xi_0^\me)^2\xi_0^\pe\right]^{1/3} t^{-1/4}\,,\nonumber\\
l_\pe&=&\sqrt{6}\left(\frac{1}{2}\frac{C\!_\pe}{C\!_\me}\right)^{2/3}
\left[\frac{\vfi_0^\me}{\vfi_0^\pe}(\xi_0^\me)^2\xi_0^\pe\right]^{1/3} t^{-1/4}\,.
\label{eq:z36-c}
\eea
At the interface we obtain
\be
\vfi(0)=\left(\frac{1}{2}\frac{C\!_\pe\xi_0^\me}{C\!_\me\xi_0^\pe}\vfi_0^\pe\right)^{2/3}(\vfi_0^\me)^{1/3}t^{1/2}\,.
\label{eq:fi36-c}
\ee

\section{Scaling considerations}
\label{sec:scaling}

Here we generalize the mean-field results obtained in the previous
section. First, we consider the order-parameter profiles in
semi-infinite systems with free and fixed boundary
conditions. These results are used afterwards to study the scaling
behavior at an interface, which separates two different semi-infinite
systems.

\subsection{Order-parameter profiles in semi-infinite systems}

We consider a semi-infinite system, which is located in the half-space $z>0$, and
which is in its bulk-ordered phase ($T \le T_c$); see in
Sec.~\ref{sec:mf-2}. As in mean-field theory, the order-parameter $\varphi(z)$ 
depends on the distance from the surface, $z$, and
approaches its bulk value, $\varphi_b \sim t^{\beta}$, for $z/\xi \gg
1$. The bulk correlation length asymptotically behaves as $\xi
\sim |t|^{-\nu}$.  These expressions generalize the mean-field results
in Eqs.~(\ref{eq:mb}) and (\ref{eq:xi}).

\subsubsection{Free boundary conditions}

At a free surface, due to the missing bonds the local order is weaker
than in the bulk. The surface order parameter displays the
so-called ordinary transition with the temperature dependence $\varphi(0)
\sim t^{\beta_1}$, where generally $\beta_1 > \beta$.  The
profile, $\varphi(z)$, which interpolates between the surface and the bulk
value has the scaling form\cite{binder83,diehl86,pleimling04}
\be
\varphi(z)=\varphi_b f_{\rm ord}\left( \frac{z+l}{\xi} \right)\;,
\label{eq:m_o}
\ee and the scaling function, $f_{\rm ord}(y)$, behaves as 
$y^{(\beta_1-\beta)/\nu}$, for $y \ll 1$. 

\subsubsection{Fixed boundary conditions}

For fixed boundary conditions, the system displays the extraordinary
surface transition and stays ordered in the surface region at the bulk
critical temperature, so that $\varphi(z)=O(1)$ as $t \to 0_\pe$ and $z\ll\xi$. This behavior
is formally equivalent to having a surface exponent, $\beta_1=0$.  The
magnetization profile can be written into an analogous form\cite{binder83,diehl86,pleimling04} as in
Eq.~(\ref{eq:m_o}):
\be
\varphi(z)=\varphi_bf_{\rm ext}\left( \frac{z+l}{\xi} \right)\;,
\label{eq:m_e}
\ee
however, now the scaling function, $f_{\rm ext}(y)$ has the asymptotic
behavior,\cite{fisher78}  $f_{\rm ext}(y) \sim y^{-\beta/\nu}$, for $y \ll 1$.

\subsection{Interface critical behavior}

Now we join the two semi-infinite systems and study the behavior of
the order-parameter in the vicinity of the interface. In general we
expect that, depending on the strength of the interface coupling, at the
bulk critical temperature the interface (i) can stay disordered for
weak couplings, which corresponds to the $\Lambda > 0$ case in
mean-field theory or (ii) can stay ordered for stronger couplings,
which is the case for $\Lambda < 0$ in mean-field theory. These two
regimes of interface criticality are expected to be separated by a
special transition point, which corresponds to $\Lambda \to \infty$ in mean-field
theory.

To construct the order-parameter profile we start with the profiles in
the semi-infinite systems and join them. First we require continuous
behavior of the profile at $z=0$, like in mean-field theory. The second
condition in mean-field theory in Eq.~(\ref{eq:bci}) cannot be directly
translated; here we just use its consequencies for the extrapolation
lengths. In the weak- and strong-coupling regimes in mean-field theory
the left-hand side of Eq.~(\ref{eq:bci}) is finite so that the derivative of the profile is
discontinuous at $z=0$ and at least one of the extrapolation lengths $l_\pme$ is
$O(1)$. The same behavior of $l_\pme$ is expected to hold in scaling theory, too.
On the other hand, at the special transition point in mean-field theory the left-hand side of
Eq.~(\ref{eq:bci}) is zero and the extrapolation lengths are divergent.
In scaling theory the asymptotic form of the extrapolation lengths is expected
to be deduced from the same condition--- i.e. from the equality of the derivatives of the profiles.
This leads to a relation  $l_\pme \sim t^{-\nu_\pme \omega_\pme}$, in which
$0 \le \omega_\pme \le1$ is defined later.

If the subsystem---say at $z>0$---has an ordinary transition
the interface magnetization follows from Eq.~(\ref{eq:m_o}) as
\be
\varphi(0) \sim t^{\beta_i},\quad \beta_i=(1-\omega_\pe)\beta_1^\pe+\omega_\pe \beta_\pe \;,
\label{eq:m_o_0}
\ee
and $\beta_\pe \le \beta_i \le \beta_1^\pe$. On the other hand, if the
subsystem---say at $z<0$---has an extraordinary transition
the interface magnetization exponent follows from Eq.~(\ref{eq:m_e}) as
\be
\beta_i=\omega_\me \beta_\me \;.
\label{eq:m_e_0}
\ee
Evidently, $\beta_i$ calculated form the two joined subsystems should have the same value.
This type of construction of the order-parameter profiles will lead to a smooth
profile at the interface provided the
extrapolation lengths are smaller or, at most, of the same order than the correlation lengths,
${\rm max}(l_\pe,l_\me) \lesssim {\rm min}(\xi_\pe,\xi_\me)$, which holds
provided
\be
{\rm max}(\omega_\pe \nu_\pe,\omega_\me\nu_\me) \leq {\rm min}(\nu_\pe,\nu_\me)\;.
\label{eq:loc}
\ee
Otherwise, the profile measured in a length scale, ${\rm min}(\xi_\pe,\xi_\me)$, has a sharp variation
at the interface and as the critical temperature is approached the profile becomes
discontinuous.
Note that in mean-field theory, with $\nu_\pe=\nu_\me$ and $\omega_\pme\leq 1$, the profile is
always smooth. 

\subsubsection{Relevance-irrelevance criterion}

Here we generalize the relevance-irrelevance criterion known to hold at
an internal defect plane with weak defect couplings.\cite{burkhardt81} If two different critical
systems are weakly coupled, the operator corresponding to the junction is the
product of the two surface magnetization operators. Consequently, its anomalous
dimension $x_i$ is given by the sum of the dimensions of the two surface operators,
$x_i=x_1^\me+x_1^\pe$, where $x_1^\pme=\beta_1^\pme/\nu_\pme$. Then the scaling
exponent of the defect, $y_i$, in a $d$-dimensional system is given by
\be
y_i=d_i-x_i=d-1-\frac{\beta_1^\pe}{\nu_\pe}-\frac{\beta_1^\me}{\nu_\me}\;.
\label{eq:relev}
\ee
where $d_i=d-1$ is the dimension of the interface.

For $y_i<0$ the weak interface coupling is irrelevant so that the defect coupling
renormalizes to zero and the defect acts as a cut in the system. Consequently the
interface critical behavior is the same as in the uncoupled semi-infinite systems
and the interface magnetization exponent is $\beta_i={\rm min}(\beta_1^\pme)$
since the stronger local order manifests itself at the interface.
In the other case, $y_i>0$, the coupling at the interface is relevant and the interface
critical behavior is expected to be controlled by a new fixed point.

For  the 2D $q$-state Potts model with $2 \le q \le 4$ we have\cite{cardy84} $x_1 \ge 1/2$; thus weak
interface coupling is expected to be irrelevant according to Eq.~(\ref{eq:relev}).

In mean-field theory, when $d$ appears in a scaling relation, it has to be replaced 
by the upper critical dimension $d_c$ for which hyperscaling is verified. 
However  we have here different values of $d_c$ for the two subsystems so that there is 
some ambiguity for the value of $d$ in Eq.~(\ref{eq:relev}). The analytical results
of Sec.~\ref{sec:mf-3} show that a weak interface coupling is also irrelevant 
in all the cases studied in mean-field theory.

\subsubsection{Weakly coupled systems}

For weak interface coupling the order parameter profile is not expected
to display a maximum at the interface. Depending on the relative
values of the critical exponents, $\beta_\pe,~ \beta_\me,~ \beta_1^\pe,~
\beta_1^\me$ it can be either a minimum, or an intermediate point of a
monotonously increasing profile. We use the same convention as in
Sec.~\ref{sec:mf-3}, that $\beta^\pe < \beta^\me$ and treat separately the
different cases.

\ssection{a}{$\beta_\pe < \beta_\me < \beta_1^\pe < \beta_1^\me$}
The order-parameter profile is obtained by joining two ordinary
surface profiles in Eq.~(\ref{eq:m_o}) both for $z<0$ and for $z>0$.
  In this case the weak coupling does not modify the asymptotic
  behavior of the more ordered, $z > 0$ subsystem. Consequently we
  have $\varphi(0) \sim \varphi_1^\pe$, $l_\pe=O(1)$ and $\omega_\pe=0$; thus
  $\beta_i=\beta_1^\pe$. From Eq.~(\ref{eq:m_o_0}) we obtain
\be
\omega_\me=\frac{\beta_1^\me-\beta_1^\pe}{\beta_1^\me-\beta_\me}\;.
\label{eq:om-m}
\ee
Note that the above reasoning leads to a smooth order-parameter profile at the interface,
if according to Eq.~(\ref{eq:loc}) we have $\omega_\me \nu_\me <
\nu_\pe$.  This type of behavior is realized in mean-field theory for
the $\varphi^4-\varphi^6$ interface for $\Lambda>0$, see in
Sec.~\ref{sec:mf-3-2}.

\ssection{b}{$\beta_\pe < \beta_\me < \beta_1^\me < \beta_1^\pe$}
In this case the profile is obtained from two ordinary subprofiles.
 The order parameter is still minimum at the interface, but it is determined
  by the $z<0$ subsystem, which has the larger surface order parameter. 
  Consequently, $\varphi(0) \sim \varphi_1^\me$, $l_\me=O(1)$ and $\omega_\me=0$; thus
  $\beta_i=\beta_1^\me$.  From Eq.~(\ref{eq:m_o_0}) we obtain
\be
\omega_\pe=\frac{\beta_1^\pe-\beta_1^\me}{\beta_1^\pe-\beta_\pe}\;.
\label{eq:om-p}
\ee
and the order-parameter profile is smooth if $\omega_\pe \nu_\pe < \nu_\me$. This
type of behavior is never realized in mean-field theory; see the
exponent relation in Eq.~(\ref{eq:scal1}).

\ssection{c}{$\beta_\pe  < \beta_1^\pe < \beta_\me < \beta_1^\me$}
In this case the order-parameter profile is monotonously increasing
  and obtained by joining an extraordinary profile in
  Eq.~(\ref{eq:m_e}) for $z<0$ with an ordinary profile in
  Eq.~(\ref{eq:m_o}) for $z>0$. The order parameter at the interface is
  determined by the surface order parameter of the $z>0$ subsystem. Then we
  have $\varphi(0) \sim \varphi_1^\pe$, $l_\pe=O(1)$ and $\omega_\pe=0$; thus
  $\beta_i=\beta_1^\pe$. From Eq.~(\ref{eq:m_o_0}) we obtain $\omega_\me=\beta_1^\pe/\beta_\me$
and the interface is smooth, provided $\omega_\me \nu_\me < \nu_\pe$. In mean-field theory this type of behavior
is realized for the $\varphi^3-\varphi^6$ interface for $\Lambda>0$; see in
Sec.~\ref{sec:mf-3-3}.

\subsubsection{Special transition point}

In this case the profile is monotonously increasing and it is constructed
by joining an extraordinary subprofile in Eq.~(\ref{eq:m_e}) for
$z<0$ with an ordinary subprofile in Eq.~(\ref{eq:m_o}) for $z>0$. As we argued
before,  the extrapolation lengths and the corresponding exponents are obtained
(i) from the continuity of the profile at $z=0$,
\be
\beta_i=\beta_\me \omega_\me=(1-\omega_\pe)\beta_1^\pe+\beta_\pe\omega_\pe\;,
\label{eq:sp1}
\ee
and (ii) from the continuity of the derivative at $z=0$, which leads to
the condition $l_\pe \sim l_\me$---consequently,
\be
\nu_i=\omega_\pe \nu_\pe = \omega_\me \nu_\me\;.
\label{eq:sp2}
\ee
The solution of Eqs.~(\ref{eq:sp1}) and (\ref{eq:sp2}) is given by
\be
\nu_i=\frac{\beta_1^\pe}{\frac{\beta_1^\pe-\beta_\pe}{\nu_\pe} +\frac{\beta_\me}{\nu_\me}}, \quad
\beta_i=\frac{\beta_\me}{\nu_\me} \nu_i\;.
\label{omega}
\ee
Let us now analyze the condition for the smooth or discontinuous nature of the interface given in Eq.~(\ref{eq:loc}).

\ssection{a}{$\nu_\me \ge \nu_\pe$}
In this case the condition is equivalent to $\beta_\me/\nu_\me > \beta_\pe/\nu_\pe$.
As we will discuss in Sec.~\ref{sec:num} this condition is satisfied in 2D for the three-  and the four-state Potts (or BW) models so that the profile is predicted to be smooth.
On the contrary for the Ising and the three-state Potts models this condition does not hold;
thus, the profile is probably sharp and becomes discontinuous at the critical temperature.
Finally, for the Ising and the BW models the relation in Eq.~(\ref{eq:loc}) is
just an equality, so that we are in a marginal situation.

\ssection{b}{$\nu_\me < \nu_\pe$}
In this case the profile is smooth, provided
\be
\beta_1^\pe < \frac{\frac{\beta_\me}{\nu_\me}-\frac{\beta_\pe}{\nu_\pe}}{\frac{1}{\nu_\me}-\frac{1}{\nu_\pe}}\;.
\ee
This type of situation seems to be less common in real systems.

\subsubsection{Strongly coupled systems}

In this case the interface stays ordered at the bulk critical temperature, so that the
profile is expected to be composed from two extraordinary subprofiles. As a consequence
the interface critical behavior is the same as in two independent semi-infinite systems, both
having an extraordinary surface transition.

\section{Numerical investigations}
\label{sec:num}

We have studied numerically the critical behavior at the interface between two
$q$-state Potts models on the square lattice, with  different values of $q$ for the two subsystems.
For a review of the Potts model, see Ref.~\onlinecite{wu82}.
In particular we considered the value $q=2$, which corresponds
the Ising model, as well as $q=3$ and $q=4$. All these systems display a
second-order phase transition for a value of the coupling given by
$e^{qK_c}=1+\sqrt{q}$, which  follows from self-duality. The associated
critical exponents are exactly known\cite{baxter82} for $q=2$ ($\beta=1/8$, $\nu=1$ and $\beta_1=1/2$)
and has been conjectured for $q=3$ ($\beta=1/9$, $\nu=5/6$ and $\beta_1=5/9$) and $q=4$  ($\beta=1/12$,
$\nu=2/3$ and $\beta_1=2/3$),  where they follow from conformal invariance\cite{cardy87} and
the Coulomb-gas mapping.\cite{nienhuis87}  We have also considered the BW
model,\cite{baxterwu} which is a triangular lattice Ising model with three-spin interactions on all the triangular faces. This model is also self-dual and  has the same critical coupling as the Ising model. This model
is exactly solved,\cite{baxter82} and it  belongs to the universality class
of the $q=4$ state Potts model, but without logarithmic corrections to scaling,
which facilitates the analysis of the numerical data.


\begin{figure}
\begin{center}
\includegraphics[width=\columnwidth]{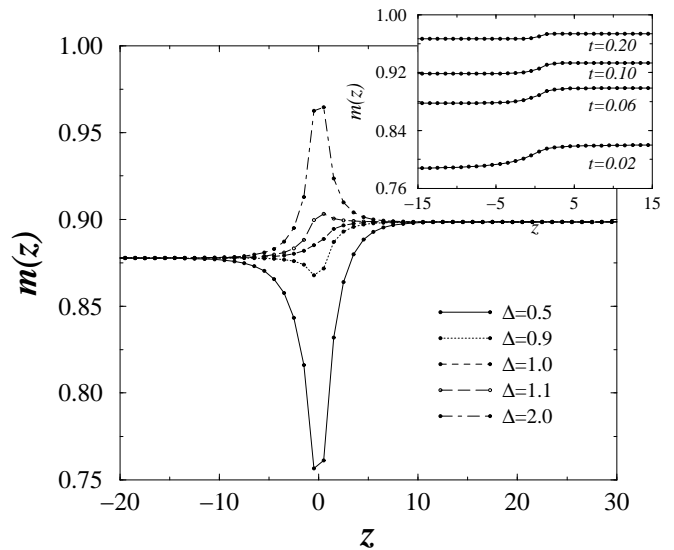}
\end{center}
\caption{Order parameter profiles of the $q=3-4$ interface at the
  reduced temperature $t=0.06$ and for different values of the
  interface coupling ratio, $\Delta=2.0,1.1,1.0,0.9$ and $0.5$ from 
  top to bottom. Inset: profiles around the
special transition point, $\Delta=1$, for different temperatures. The results indicate a
smooth profile at the transition point.}
\label{fig4}
\end{figure}


We have performed Monte Carlo simulations on 2D systems consisting of
two $L \times L$ subsystems which interact directly through interface couplings $K_i$ (between adjacent spins of the two subsystems) such that $K_i/K_\pe=\Delta$.
 Here $K_\pe$ is the coupling in the half-space $z>0$, which corresponds to the subsystem having the larger
value of $q$, thus the larger magnetization. Periodic boundary conditions are applied
in both directions. Using the Swendsen-Wang cluster-flip algorithm\cite{sw} we have calculated the magnetization
profile in systems with size up to $L=300$ for different values of the reduced temperature  $t=(T_c-T)/T_c$
and  coupling ratio $\Delta$. Depending on the size of the system and the temperature we have skipped the
first $5-20 \times 10^4$ thermalization steps and the thermal averages were taken  over $6-20 \times 10^6$
MC steps. We have checked that the magnetization profiles, at the
reduced temperatures we used,  does not show any noticeable finite-size effects. From the magnetization data at
the interface we have calculated effective, temperature-dependent interface exponents given by
\be
\beta_i(t)=\frac{\log[ m(t+\delta)/m(t-\delta)]}{\log [(t+\delta)/(t-\delta)]}\;,
\label{eq:beta_eff}
\ee
which approach the true exponents as $\delta \to 0$ and $t \to 0$. For the  Ising-BW
interface,  we have also made calculations at the critical temperature in order to  check
the finite-size scaling properties of the profiles. In the following we
present the numerical results for the $q=2-3$, $q=3-4$ and Ising-BW
interfaces. In each case we have a different type of special transition, separating the ordinary
and the extraordinary transition regimes.

\subsection{$\bm{q=3-4}$ interface}


\begin{figure}
\begin{center}
\includegraphics[width=\columnwidth]{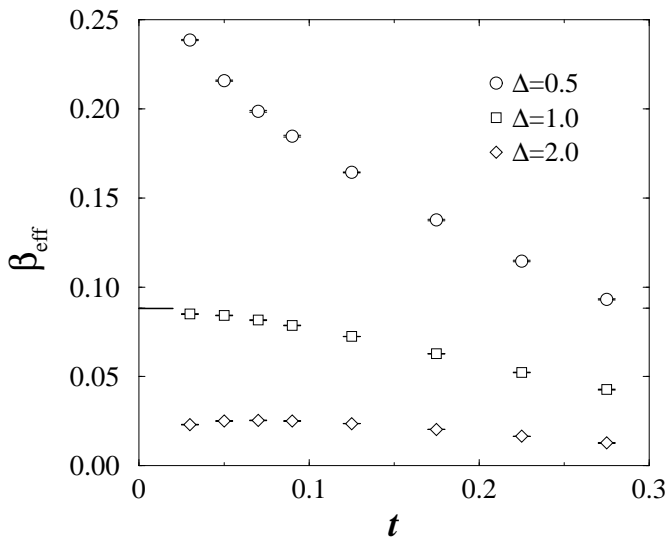}
\end{center}
\caption{Effective magnetization exponents measured on the two sides
  of the $ q=3-4$ interface for three different values of the
  coupling ratio, $\Delta$. At the special transition point, $\Delta\approx 1$, the
  theoretical prediction from Eq.~(\ref{omega}) is indicated by a bar.
  }
\label{fig5}
\end{figure}


We start in Fig.~\ref{fig4} with a  presentation of  the order-parameter profiles, in the vicinity of  the critical temperature,
for different values of the interface coupling.
Here one can differentiate between the ordinary transition regime for small $\Delta$, in which
the magnetization at the interface vanishes faster than in  the bulk of the two subsystems, and the extraordinary
transition regime for large $\Delta$,  where the interface magnetization keeps a finite value.
The special transition separating these two regimes is located at $\Delta \approx 1$. The inset of
Fig.~\ref{fig4} shows the evolution of the interface at  the special transition point
as the bulk transition point is approached. Here the criterion in Eq.~(\ref{eq:loc}) is satisfied,
since, as discussed below Eq.~(\ref{omega}), $\beta_\me/\nu_\me=2/15 > \beta_\pe/\nu_\pe=1/8$. Thus
the profile is predicted to be smooth, which is in accordance with the numerical results.

The values of the effective, temperature-dependent exponents,
as defined in Eq.~(\ref{eq:beta_eff}), are presented in Fig.~\ref{fig5}
for three values of $\Delta$, corresponding to the different transition
regimes. Clearly the values of  the effective exponents are affected by strong crossover  effects for small
$t$, since the limiting values are $\beta_i=\beta_1(q=3)=5/9$ for $\Delta <1$ and $\beta_i=0$ for $\Delta >1$,  according to scaling theory. Unfortunately, due to finite-size effects we could not go closer to the critical point.
At the special transition point, however, the crossover effects are weaker and the
effective exponents are close to the theoretical prediction in Eq.~(\ref{omega}), $\beta_i=32/363=0.088$.

\subsection{$\bm{q=2-3}$ interface}


\begin{figure}
\begin{center}
\includegraphics[width=\columnwidth]{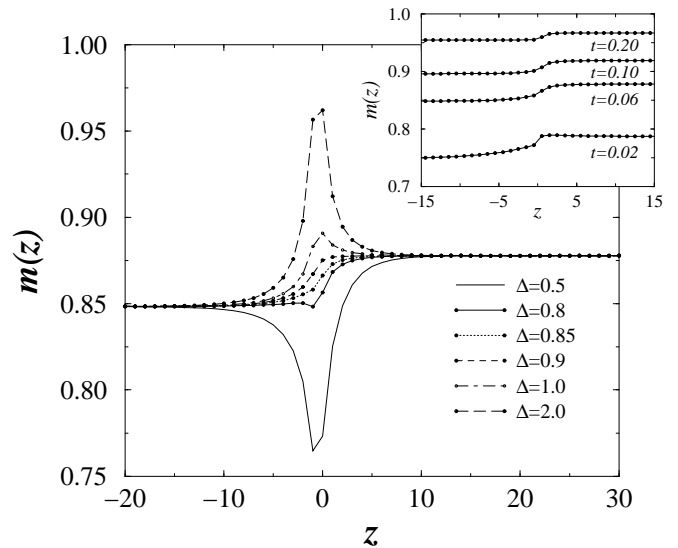}
\end{center}
\caption{The same as in Fig.~\ref{fig4} for the $q=2-3$ interface with
  interface coupling ratio $\Delta=2.0,1.0,0.9,0.85,0.8$ and $0.5$ from
  top to bottom and $t=0.06$. Inset: profiles around the
special transition point, $\Delta_c=0.85$, for different temperatures. The results indicate that the profile becomes discontinuous at the critical temperature.}
\label{fig6}
\end{figure}



\begin{figure}
\begin{center}
\includegraphics[width=\columnwidth]{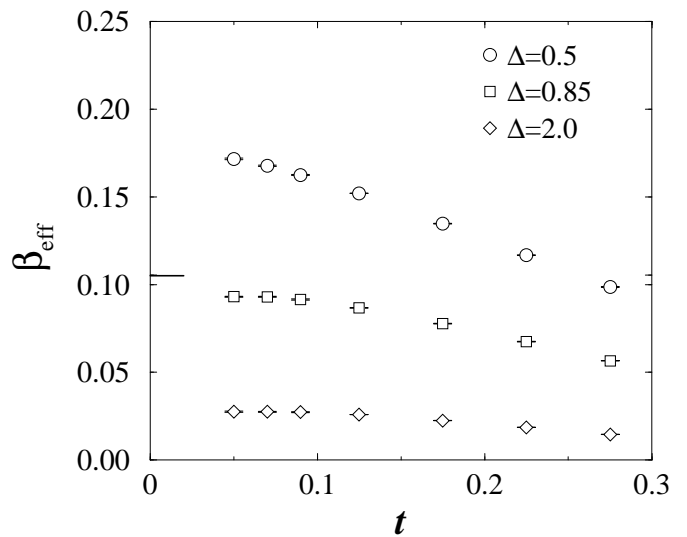}
\end{center}
\caption{The same as in Fig.~\ref{fig5} for the $q=2-3$ interface.}
\label{fig7}
\end{figure}


We have performed a similar investigation for the interface critical behavior of
the $q=2-3$ system and the results are summarized in Figs.~\ref{fig6}
and~\ref{fig7}. Here one can also identify the ordinary and the extraordinary transition
regimes (see Fig.~\ref{fig6}), which are separated by the special transition
around $\Delta \approx 0.85$. However, as can be seen in the inset of Fig.~\ref{fig6}, the behavior around
the special transition point is more complex  than for the $q=2-3$ interface.
The evolution of the profile suggests  the existence of a discontinuity at the transition
temperature. This is in accordance with the scaling criterion in Eq.~(\ref{eq:loc}) since,
as discussed below Eq.~(\ref{omega}), $\beta_\me/\nu_\me=1/8 < \beta_\pe/\nu_\pe=2/15$ leads to a
discontinuous profile. Due to this discontinuity, it is more difficult to locate precisely  the special transition point and to determine the associated  interface exponent $\beta_i$. The measured
effective interface exponents are shown in  Fig.~\ref{fig7} for three values of $\Delta$ corresponding
to the different interface fixed points. The crossover  effects are strong
but, at the special transition point, our estimates are compatible with the scaling
prediction in Eq.~(\ref{omega}), $\beta_i=25/237=0.105$.

\subsection{Ising-BW interface}


\begin{figure}
\begin{center}
\includegraphics[width=\columnwidth]{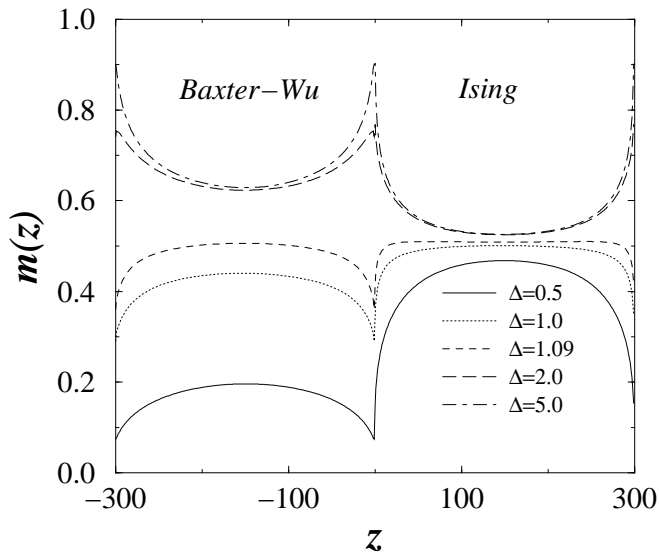}
\end{center}
\caption{Critical magnetization profiles in the Ising-BW system  with 
 two symmetrically placed interfaces  for different values of the coupling ratio $\Delta$.}
\label{fig8}
\end{figure}



\begin{figure}
\begin{center}
\includegraphics[width=\columnwidth]{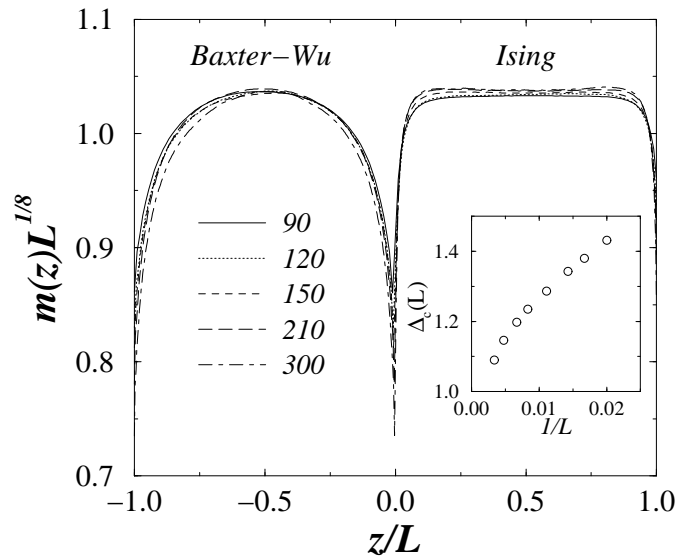}
\end{center}
\caption{Scaled magnetization profiles for the coupled  Ising-BW
systems at the common critical temperature for $L=90$ --- $300$. The interface
coupling ratio is fixed at the critical value, $\Delta_c(L)$, for which the two maxima of the curves are identical.
The inset  gives the effective critical interface coupling  as a function of the inverse size.}
\label{fig9}
\end{figure}


The interface between the Ising model and the BW model (or the four-state Potts model)
has some special features. These are mainly due to the fact that the anomalous dimension of
the bulk magnetization in the two systems has the same value, $\beta_\me/\nu_\me = \beta_\pe/\nu_\pe=1/8$. 
Consequently one can
define and numerically study the finite-size scaling properties of the magnetization profile at
the phase-transition point, since it is expected to scale as  $m(z,L)=L^{-1/8}f(z/L)$.
The scaling function $f(y)$ is expected to depend on the value of the interface coupling ratio
$\Delta$ and we have studied this quantity numerically. 

The magnetization profiles at the critical temperature for different values of the interface
coupling ratio $\Delta$ are given in Fig.~\ref{fig8}. It is interesting to notice that the shape
of the curves as well as the relative heights of the profiles in the two subsystems vary with
the interface coupling. For $\Delta<\Delta_c \approx 1$, the interface stays disordered and
the interface critical behavior is governed by the surface exponent of the Ising model. The
larger bulk value on the Ising side is understandable since the profile on the right side
is more singular, $\beta_1^\me/\nu_\me > \beta_1^\pe/\nu_\pe$; see below Eq.(\ref{eq:m_o}).
On the contrary, for $\Delta>\Delta_c$ the interface is ordered at the bulk critical temperature
and the profiles decay towards the bulk values.

At the special transition point $\Delta=\Delta_c$, the profile has a universal form in terms of  $L^{1/8}m(z/L)$. This is illustrated in Fig.~\ref{fig9}, in which for each finite
system a critical value $\Delta_c(L)$ is calculated from the condition that the two maxima of the curves have identical values and the profile is measured at that interface coupling. The size-dependent effective interface coupling ratio, $\Delta_c(L)$, shown in the inset of Fig.~\ref{fig9},
seems to tend to a limiting value,  $\Delta_c \approx 1$. The scaled magnetization profiles have different
characteristics in the two subsystems. In the BW model, having the smaller correlation length, the profile has a smooth variation. On the contrary,  on the Ising side,  the profile has a quasidiscontinuous
nature at the interface, which is probably related to the fact that in the criterion of Eq.~(\ref{eq:loc})
the equality holds.

\section{Discussion}
\label{sec:disc}

In this paper we have studied the critical behavior at the interface between two subsystems displaying 
a second-order phase transition. We assumed that the critical temperatures are identical but the sets of critical
exponents (i.e., the universality classes  of the transitions) are different for the two subsystems. By varying the interface couplings, we monitored
the  order at the interface and studied the behavior of the order-parameter
profile as the critical temperature is approached. We provided a detailed analytical solution of
the problem in the framework  of  mean-field theory, which leads to a physical picture
which is useful for the study of realistic systems. Solutions  of the mean-field equations are obtained by
adjusting the order-parameter profiles of the two semi-infinite subsystems through the introduction of appropriate extrapolation
lengths on the two sides. The same strategy has been applied  in the frame of a phenomenological
scaling approach. As a result, basically three types of interface critical behavior are  observed.
For weak interface couplings the interface renormalizes to an effective cut and we are left with
the surface critical behavior of the subsystems. In the limit of strong interface couplings,
the renormalization leads to infinitely strong local couplings and thus interface order at the bulk
critical point. Finally, for some intermediate value of the interface couplings, the interface displays a
special transition, which is characterized by a new critical exponent for the order parameter in the interface region.
In the scaling theory this exponent can be expressed in terms of the bulk and surface exponents of the semi-infinite subsystems.  

These  results have been  tested through  large scale Monte Carlo simulations, in which
the critical behavior at the interface between 2D Ising, Potts and BW models was studied and satisfactory agreement has been found. However, it would be interesting to confirm the analytical expressions for the interface exponents
through a field-theoretical renormalization group study, using the methods of Ref.~\onlinecite{diehl86}.

The results obtained in this paper can be generalized into different directions. First we mention the case
when the critical temperatures of the subsystems are not exactly equal but differ by an amount, $\Delta T_c$. If the deviation in temperature from the average value, $\overline{T}_c=(T_c^\me+T_c^\pe)/2$, is small but
satisfies $\overline{T}_c-T \gg \Delta T_c$, then our results are still valid. Our second remark concerns 3D systems in which sufficiently enhanced interface couplings may lead to an independent ordering of the interface above the bulk critical temperatures. In semi-infinite systems this phenomena is called the {\it surface transition}.\cite{binder83,diehl86,pleimling04} At the bulk critical temperature the ordered interface then shows a singularity, which is analogous to the {\it extraordinary transition} in semi-infinite systems. The singularities at the {\it interface} and {\it extraordinary interface transitions} remain to be determined, even in the mean-field approach. Third, we can mention that non-trivial interface critical behavior could be observed when one of the subsystems displays a first-order
transition. It is known for semi-infinite systems that the surface may undergo a continuous transition,
which, however, has an anisotropic scaling character, even if the bulk transition is discontinuous.\cite{lipowsky84,ti02}
Similar phenomena can happen at an interface, too. Our final remark concerns the localization-delocalization
transition of the interface provided an external ordering field is applied. For two subsystems having
the same $\varphi^4$ mean-field theory and the same\cite{sevrin89} or different\cite{igloi90} critical temperatures,  this wetting problem has already been solved. This solution could be generalized for
subsystems having different field-theoretical descriptions.

\begin{acknowledgments}
  F.\'A.B. thanks the Minist\`ere Fran\c{c}ais des
  Affaires \'Etrang\`eres for a research grant. This work has been
  supported by the Hungarian National Research Fund under Grant Nos.
  OTKA TO34183, TO37323, TO48721, MO45596 and M36803. Some 
  simulations have been performed at CINES Montpellier under Project
  No. pnm2318.  The Laboratoire de Physique des Mat\'eriaux is Unit\'e
  Mixte de Recherche CNRS No. 7556.
\end{acknowledgments}

\appendix

\section{Surface critical behavior}
\label{app:surf}
The calculation of the surface behavior is straightforward when $l/\xi\to0$ at the critical point.
Here we give some details about the two cases where the surface boundary condition leads to a constant value for $l/\xi$. 

\subsection{$\bm{\vfi^3}$ model with $\bm{\Lambda>0}$}
\label{app:surf-1}
Using the results of Eq.~(\ref{eq:fi3-b}), the boundary condition in~(\ref{eq:bcz}) is rewritten as
\be
\frac{C_s}{2\Lambda}\!\left[3\tanh^2\!\left(\!\frac{l}{2\xi}\!\right)\!\!-\!\!1\right]\!\!
=\!\frac{3C}{2\xi}\sinh\!\left(\!\frac{l}{2\xi}\!\right)\!\cosh\!^{-3}\!\left(\!\frac{l}{2\xi}\!\right).
\label{eq:bc3-1}
\ee
Let
\be 
3\tanh^2\left(\frac{l}{2\xi}\right)-1=\alpha t^{1/2}\,,\quad \hat{\vfi}(0)=\frac{\alpha}{2}t^{1/2}\,.
\label{eq:3alpha-1}
\ee
Then, to leading order
\be
\tanh\left(\frac{l}{2\xi}\right)=\frac{1}{\sqrt{3}}\,,\quad
\cosh^{-2}\left(\frac{l}{2\xi}\right)=\frac{2}{3}\,.
\label{eq:3thch}
\ee
The first relation gives $l$ in Eq.~(\ref{eq:l3-b}), and~(\ref{eq:bc3-1}) leads to
\be
\alpha=\frac{2C\Lambda}{\sqrt{3}C_s\xi_0}
\label{eq:3alpha-2}
\ee
Finally, combining Eqs.~(\ref{eq:3alpha-1}) and~(\ref{eq:3alpha-2}), one obtains the value of $\vfi(0)$ given in Eq.~(\ref{eq:f3-b}).

\subsection{$\bm{\vfi^6}$ model with $\bm{\Lambda<0}$}
\label{app:surf-2}
Since the surface is ordered, the profile is given by Eq.~(\ref{eq:fi3-b}). The boundary condition in~(\ref{eq:bcz}) translates into
\bea
&&\frac{\sqrt{2}C_s}{\vert\Lambda\vert}\left[3\tanh^2\left(\frac{l}{2\xi}\right)-1\right]^{-1/2}\!\!\!=\frac{3\sqrt{2}C}{2\xi}\sinh\left(\frac{l}{2\xi}\right)\nonumber\\
&&\ \ \ \times
\cosh^{-3}\left(\frac{l}{2\xi}\right)\left[3\tanh^2\left(\frac{l}{2\xi}\right)-1\right]^{-3/2}\!\!\!.
\label{eq:bc6-1}
\eea
The boundary condition is satisfied when
\be
\sqrt{3\tanh^2\left(\frac{l}{2\xi}\right)-1}=\alpha t^{1/4}\,,
\label{eq:6alpha-1}
\ee
so that:
\be
\hat{\vfi}(0)=\frac{\sqrt{2}}{\alpha}t^{-1/4}\,.
\label{eq:6fi0}
\ee
Eq.~(\ref{eq:6alpha-1}) gives the value of $l$ in Eq.~(\ref{eq:l6-a}). Using the values given in~(\ref{eq:3thch}) which remain valid here together with Eq.~(\ref{eq:6alpha-1}) in~(\ref{eq:bc6-1}), one obtains:
\be
\alpha=\left(\frac{2C\vert\Lambda\vert}{\sqrt{3}C_s\xi_0}\right)^{1/2}\,.
\label{eq:6alpha-2}
\ee
Inserting this expression in~(\ref{eq:6fi0}) leads to the surface order parameter given in~Eq.~(\ref{eq:f6-a}).

\section{Interface critical behavior}
\label{app:inter}
In this Appendix we give some details about the calculations of $l_\pme$ and $\vfi(0)$ limiting ourselves to three representative cases. Other results are easily obtained using similar methods.

\subsection{$\bm{l_\pme/\xi_\pme\ll1}$}
\label{app:inter-1}

This situation is encountered for the $\vfi^3-\vfi^4$ interface with $\Lambda<0$ and $\Lambda\to\infty$ as well as for the $\vfi^4-\vfi^6$ and $\vfi^3-\vfi^6$ interfaces with $\Lambda>0$ and $\Lambda\to\infty$. 
Here we consider as an example the $\vfi^3-\vfi^4$ interface with $\Lambda<0$.

The boundary conditions in Eq.~(\ref{eq:bci}) are satisfied with $\vfi_\me(z)$ and $\vfi_\pe(z)$ given by~(\ref{eq:fi3-a}) and~(\ref{eq:fi4-a}) and reads
\bea
&&\!\!\!\!\vfi(0)\!=\vfi_0^\pe t^{1/2}\coth\left(\!\frac{l_\pe}{2\xi_\pe}\!\right)\!=\!\frac{\vfi_0^\me t}{2}\left[3\coth^2\!\left(\!\frac{l_\me}{2\xi_\me}\!\right)\!-\!1\right],\nonumber\\
&&\!\!\!\!\!\!\frac{C\!_\pe\vfi_0^\pe t^{1/2}}{2\xi_\pe}\!\sinh^{-2}\!\left(\!\frac{l_\pe}{2\xi_\pe}\!\!\right)
\!+\!\frac{3C\!_\me\vfi_0^\me t}{2\xi_\me}\cosh\!\left(\!\frac{l_\me}{2\xi_\me}\!\!\right)
\!\sinh^{-3}\!\left(\!\frac{l_\me}{2\xi_\me}\!\!\right)\nonumber\\
&&\ \ \ \ =\frac{C_i \vfi_0^\pe t^{1/2}}{\vert\Lambda\vert}\coth\left(\frac{l_\pe}{2\xi_\pe}\right)\,.
\label{eq:bc34-1}
\eea
With $l_\pme/\xi_\pme\ll1$, one may expand the hyperbolic functions in powers of $l_\pme/(2\xi_\pme)$
To leading order, the first equation in~(\ref{eq:bc34-1}) gives
\be
\vfi(0)=\frac{3\vfi_0^\me t}{2}\left(\frac{2\xi_\me}{l_\me}\right)^2
=\frac{2\vfi_0^\pe\xi_0^\pe}{l_\pe}\,,
\label{eq:vfi34-1}
\ee
so that
\be
\frac{l_\me}{2\xi_\me}=\sqrt{\frac{3\vfi_0^\me l_\pe t}{4\vfi_0^\pe\xi_0^\pe}}\,.
\label{eq:zm34-1}
\ee
Introducing this result in the second equation, one obtains an equation of the second degree in $x=\sqrt{l_\pe}$:
\be
-x^2+2\frac{C\!_\me\vert\Lambda\vert}{C_i\xi_0^\me}\sqrt{\frac{\vfi_0^\pe\xi_0^\pe}{3\vfi_0^\me}}x
+\frac{C\!_\pe\vert\Lambda\vert}{C_i}=0\,.
\label{eq:zpx34-1}
\ee
Thus we have
\bea
\sqrt{l_\pe}&=&f\frac{C\!_\me\vert\Lambda\vert}{C_i\xi_0^\me}\sqrt{\frac{\vfi_0^\pe\xi_0^\pe}{3\vfi_0^\me}}\,,\nonumber\\
f&=&\sqrt{1+3\frac{C\!_\pe C_i\vfi_0^\me(\xi_0^\me)^2}{C_\me^2\vfi_0^\pe\xi_0^\pe\vert\Lambda\vert}}\,.
\label{eq:zp34-1}
\eea
This last result together with Eqs.~(\ref{eq:zm34-1}) and~(\ref{eq:vfi34-1}) leads to the expressions given in~(\ref{eq:z34-a}) 
and~(\ref{eq:fi34-a}).

\subsection{$\bm{l_\me/\xi_\me}\to$ const., $\bm{l_\pe/\xi_\pe\ll1}$}
\label{app:inter-2}

This behavior is obtained only for the $\vfi^3-\vfi^4$ interface with $\Lambda>0$. When $\Lambda$ is smaller than a critical value $\Lambda_c$ to be determined later, the profiles $\vfi_\me(z)$ and $\vfi_\pe(z)$ are given by~(\ref{eq:fi3-b}) and~(\ref{eq:fi4-b}). They lead to the following boundary conditions:
\bea
&&\!\!\!\vfi(0)\!=\!\vfi_0^\pe t^{1/2}\tanh\!\left(\!\frac{l_\pe}{2\xi_\pe}\!\right)\!=\!\frac{\vfi_0^\me t}{2}\!\left[3\tanh^2\!\left(\!\frac{l_\me}{2\xi_\me}\!\right)\!-\!1\right],\nonumber\\
&&\!\!\!\!\!\!\frac{C\!_\pe\vfi_0^\pe t^{1/2}}{2\xi_\pe}\!\cosh^{-2}\!\left(\!\frac{l_\pe}{2\xi_\pe}\!\!\right)
\!+\!\frac{3C\!_\me\vfi_0^\me t}{2\xi_\me}\!\sinh\!\left(\!\frac{l_\me}{2\xi_\me}\!\!\right)
\!\cosh^{-3}\!\!\left(\!\frac{l_\me}{2\xi_\me}\!\!\right)\nonumber\\
&&\ \ \ \ =\frac{C_i \vfi_0^\pe t^{1/2}}{\Lambda}\tanh\!\left(\!\frac{l_\pe}{2\xi_\pe}\!\!\right)\,.
\label{eq:bc34-2}
\eea
With 
\be
\frac{l_\me}{2\xi_\me}=\ksm\,,
\label{eq:ksm}
\ee
the first equation in~(\ref{eq:bc34-2}) gives
\be
\vfi(0)=\vfi_0^\pe t\left(\frac{l_\pe}{2\xi_0^\pe}\right)=\frac{\vfi_0^\me t}{2}(3\tanh^2\ksm-1)\,.
\label{eq:vfi34-2}
\ee
It follows that
\be
l_\pe=\frac{\vfi_0^\me}{\vfi_0^\pe}(3\tanh^2\ksm-1)\xi_0^\pe\,.
\label{eq:a34}
\ee
The second equation in~(\ref{eq:bc34-2}) can be rewritten as
\be
C\!_\pe\vfi_0^\pe
+3C\!_\me\vfi_0^\me\frac{\xi_0^\pe\sinh \ksm}{\xi_0^\me\cosh^3\ksm} t^{1/2}
=C_i \vfi_0^\pe \frac{l_\pe}{\Lambda}\,.
\label{eq:b34}
\ee
Close to the critical point, the second term can be neglected so that
\be
l_\pe=\frac{C\!_\pe}{C_i}\Lambda\,.
\label{eq:zp34-2}
\ee
Combining this result with~(\ref{eq:a34}), one obtains
\be
\tanh \ksm=\sqrt{\frac{1+2\Lambda/\Lambda_c}{3}}\,,\qquad
\Lambda_c=2\frac{C_i \vfi_0^\me}{C\!_\pe\vfi_0^\pe}\xi_0^\pe\,.
\label{eq:th}
\ee
Since $\tanh \ksm\leq1$, this solution remains acceptable as long as 
$\Lambda\leq\Lambda_c$. Eqs.~(\ref{eq:ksm}), 
(\ref{eq:vfi34-2}), (\ref{eq:zp34-2}) and~(\ref{eq:th}) immediately lead to 
the expressions given in~(\ref{eq:z34-b1}) and (\ref{eq:fi34-b}).

When $\Lambda=\Lambda_c$, $l_\me$ diverges and the order parameter remains constant, keeping its bulk value 
on the $\vfi^3$ side of the interface. 

When $\Lambda>\Lambda_c$, the profile is always increasing. Then $\vfi_\me(z)$ is given by Eq.~(\ref{eq:fi3-a}) and the boundary conditions are changed into
\bea
&&\!\!\!\vfi(0)\!=\!\vfi_0^\pe t^{1/2}\tanh\!\left(\!\frac{l_\pe}{2\xi_\pe}\!\!\right)\!=\!\frac{\vfi_0^\me t}{2}\!\left[3\coth^2\!\left(\!\frac{l_\me}{2\xi_\me}\!\!\right)\!-\!1\right],\nonumber\\
&&\!\!\!\!\!\!\frac{C\!_\pe\vfi_0^\pe t^{1/2}}{2\xi_\pe}\cosh^{-2}\!\left(\!\frac{l_\pe}{2\xi_\pe}\!\!\right)
\!-\!\frac{3C\!_\me\vfi_0^\me t}{2\xi_\me}\cosh\!\left(\!\frac{l_\me}{2\xi_\me}\!\!\right)
\!\sinh^{-3}\!\left(\!\frac{l_\me}{2\xi_\me}\!\!\right)\nonumber\\
&&\ \ \ \ =\frac{C_i \vfi_0^\pe t^{1/2}}{\Lambda}\tanh\!\left(\!\frac{l_\pe}{2\xi_\pe}\!\right)\,.
\label{eq:bc34-3}
\eea
Inserting
\be
\frac{l_\me}{2\xi_\me}=\kgr
\label{eq:kgr}
\ee
into the first equation of~(\ref{eq:bc34-2}) leads to
\be
\vfi(0)=\vfi_0^\pe t\left(\frac{l_\pe}{2\xi_0^\pe}\right)=\frac{\vfi_0^\me t}{2}(3\coth^2\kgr-1)
\label{eq:vfi34-3}
\ee
and
\be
l_\pe=\frac{\vfi_0^\me}{\vfi_0^\pe}(3\coth^2\kgr-1)\xi_0^\pe\,.
\label{eq:c34}
\ee
From the second equation in~(\ref{eq:bc34-3}) one deduces
\be
C\!_\pe\vfi_0^\pe
-3C\!_\me\vfi_0^\me\frac{\xi_0^\pe\cosh \kgr}{\xi_0^\me\sinh^3\kgr} t^{1/2}
=C_i \vfi_0^\pe \frac{l_\pe}{\Lambda}\,,
\label{eq:d34}
\ee
where the second term can be neglected close to the critical point. Thus $l_\pe$ is still given by
\be
l_\pe=\frac{C\!_\pe}{C_i}\Lambda\,.
\label{eq:zp34-3}
\ee
Comparing with~(\ref{eq:c34}), one obtains
\be
\coth \kgr=\sqrt{\frac{1+2\Lambda/\Lambda_c}{3}}\,,
\label{eq:coth}
\ee
with the value of $\Lambda_c$ given in Eq.~(\ref{eq:th}).
Since $\coth \kgr\geq1$, this new solution replaces the preceding one when
$\Lambda\geq\Lambda_c$. The results given in~(\ref{eq:z34-b2}) and (\ref{eq:fi34-b}) follow from Eqs.~(\ref{eq:kgr}), 
(\ref{eq:vfi34-3}), (\ref{eq:zp34-3}) and~(\ref{eq:coth}).

\subsection{$\bm{l_\me/\xi_\me\ll1}$, $\bm{l_\pe/\xi_\pe\to}$ const.}
\label{app:inter-3}

This is the situation encountered for the $\vfi^4-\vfi^6$ and the $\vfi^3-\vfi^6$ interfaces with $\Lambda<0$. The treatment is similar in both cases but we give some details for the $\vfi^3-\vfi^6$ interface which is a little more complicated. The interface is more ordered than the bulk so that the profiles are given by~(\ref{eq:fi3-a}) for $z<0$ and~(\ref{eq:fi6-a}) for $z>0$. They lead to the following boundary conditions:
\bea
\!\!\!\!\vfi(0)&=&\sqrt{2}\vfi_0^\pe t^{1/4}\left[3\tanh^2
\left(\frac{l_\pe}{2\xi_\pe}\right)-1\right]^{-1/2}\nonumber\\
&=&\frac{\vfi_0^\me t}{2}
\left[3\coth^2\left(\frac{l_\me}{2\xi_\me}\right)-1\right],\nonumber\\
&&\!\!\!\!\!\!\!\!\!\!\!\!\!\!\!\!\!\!\!\!\!\!\!\!\!\!\!\!\!\frac{3\sqrt{2}C\!_\pe\vfi_0^\pe t^{1/4}}{2\xi_\pe}\sinh\left(\frac{l_\pe}{2\xi_\pe}\right)\cosh^{-3}
\left(\frac{l_\pe}{2\xi_\pe}\right)\nonumber\\
&\times&\left[3\tanh^2\left(\frac{l_\pe}{2\xi_\pe}\right)-1\right]^{-3/2}\nonumber\\
&+&\frac{3C\!_\me\vfi_0^\me t}{2\xi_\me}\cosh\left(\frac{l_\me}{2\xi_\me}\right)
\sinh^{-3}\left(\frac{l_\me}{2\xi_\me}\right)\nonumber\\
&=&\frac{\sqrt{2}C_i \vfi_0^\pe t^{1/4}}{\vert\Lambda\vert}
\left[3\tanh^2\left(\frac{ l_\pe}{2\xi_\pe}\right)-1\right]^{-1/2}\!\!\!\!\!\!\!\!\!\!\!\!.
\label{eq:bc46-1}
\eea
As in Appendix~\ref{app:surf-2}, the solution is obtained by assuming that close to the critical point:
\be
\sqrt{3\tanh^2\left(\frac{ l_\pe}{2\xi_\pe}\right)-1}=\alpha t^{1/4}\,.
\label{eq:alpha}
\ee
From this expression one deduces the leading contribution to $l_\pe$ given in~(\ref{eq:z36-a}).

With $l_\me/\xi_\me\ll1$ the first equation in~(\ref{eq:bc46-1}) can be rewritten as
\be
\vfi(0)=\frac{\sqrt{2}\vfi_0^\pe}{\alpha}=6\vfi_0^\me\left(\frac{\xi_0^\me}{l_\me}\right)^2\,.
\label{eq:vfi36-1}
\ee
Thus we have
\be
l_\me=\xi_0^\me\sqrt{3\sqrt{2}\alpha\frac{\vfi_0^\me}{\vfi_0^\pe}}\,.
\label{eq:zm36-1}
\ee
The second equation in~(\ref{eq:bc46-1}) allows us to determine the value of $\alpha$. Actually, we obtain the following equation for $x=1/\sqrt{\alpha}$:
\bea
x^4+ax-b&=&0\,,\nonumber\\
a=2^{3/4}\frac{C\!_\me\xi_0^\pe}{C\!_\pe\xi_0^\me}\sqrt{\frac{\vfi_0^\pe}{\vfi_0^\me}}\,,
&& b=\sqrt{3}\frac{C_i\xi_0^\pe}{C\!_\pe\vert\Lambda\vert}\,.
\label{eq:x36-1}
\eea
It is easy to verify that this equation has a single real positive root, $x_0$ . Below, we evaluate $x_0$ in the two limiting cases, $x_0\ll1$ and  $x_0\gg1$.

When $x_0\ll1$, one can iterate the relation 
\be
x_0= \frac{b-x_0^4}{a}=\frac{b}{a}\left(1-\frac{x_0^4}{b}\right)\,,
\label{eq:iter-1}
\ee
following from Eq.~(\ref{eq:x36-1}). We obtain
\be
x_0=\frac{b}{a}\left[1+O\left(\frac{b^3}{a^4}\right)\right]
\simeq\frac{\sqrt{3}C_i\xi_0^\me}{2^{3/4}C\!_\me\vert\Lambda\vert}\sqrt{\frac{\vfi_0^\me}{\vfi_0^\pe}}
=\alpha^{-1/2}\,.
\label{eq:x0-1}
\ee
This result is valid as long as $b^3\ll a^4$---i.e., when
\be
\vert\Lambda\vert\gg\Lambda^*=C_i
\left(\frac{\vfi_0^\me}{\vfi_0^\pe}\right)^{2/3}
\left(\frac{C\!_\pe }{\xi_0^\pe}\right)^{1/3}
\left(\frac{\xi_0^\me}{C\!_\me}\right)^{4/3}\,.
\label{eq:cross-1}
\ee
Combining~(\ref{eq:vfi36-1}), (\ref{eq:zm36-1}) and (\ref{eq:x0-1}) one easily obtains the results given in 
Eqs.~(\ref{eq:z36-a}) and (\ref{eq:fi36-a}) for $\vert\Lambda\vert\gg\Lambda^*$.

The relation following from Eq.~(\ref{eq:x36-1}) which is appropriate when $x_0\gg1$ is
\be
x_0= b^{1/4}\left(1-\frac{ax_0}{b}\right)^{1/4}\simeq b^{1/4}\left(1-\frac{ax_0}{4b}\right)\,,
\label{eq:iter-2}
\ee
so that
\be
x_0= b^{1/4}\left[1+O\left(\frac{a}{b^{3/4}}\right)\right]
\simeq\left(\sqrt{3}\frac{C_i\xi_0^\pe}{C\!_\pe\vert\Lambda\vert}\right)^{1/4}=\alpha^{-1/2}\,.
\label{eq:x0-2}
\ee
The correction term can be neglected when $a^4\ll b^3$---i.e., when $\vert\Lambda\vert\ll\Lambda^*$.
Then Eqs.~(\ref{eq:vfi36-1}), (\ref{eq:zm36-1}) together with~(\ref{eq:x0-2}) lead to the results 
given in~(\ref{eq:z36-a}) and (\ref{eq:fi36-a}) for $\vert\Lambda\vert\ll\Lambda^*$.

\end{document}